\documentclass[a4paper]{cas-sc}

\usepackage[numbers,sort&compress]{natbib}
\usepackage{bbding}
\usepackage{amsmath,amssymb}
\usepackage{booktabs}
\usepackage{siunitx}
\usepackage{threeparttable}
\usepackage{array}
\usepackage{multirow}
\usepackage{enumitem}
\usepackage{textcomp}
\usepackage{framed}    
\usepackage{needspace}
\usepackage{multicol} 
\usepackage{afterpage}
\newcommand{\printNomenclatureTopNextPage}{%
\afterpage{%
\noindent\makebox[\textwidth][c]{%
\begin{minipage}{0.98\textwidth}
\small
\begin{framed}
\setlength{\columnsep}{1.0cm}
\begin{multicols}{2}
  \printnomenclature[1.6cm]
\end{multicols}
\end{framed}
\end{minipage}
}
\vspace{2mm}
}%
}

\usepackage[official]{eurosym}
\usepackage{nomencl}
\makenomenclature

\definecolor{linkblue}{RGB}{0, 127, 172} 

\hypersetup{
    colorlinks=true,
    linkcolor=linkblue,   
    citecolor=linkblue,   
    urlcolor=linkblue    
}

\renewcommand\nomgroup[1]{%
  \item[\bfseries
  \ifstrequal{#1}{A}{Abbreviation}{%
  \ifstrequal{#1}{P}{Subscript}{%
  \ifstrequal{#1}{S}{Symbols}{%
  \ifstrequal{#1}{V}{Variables}{}}}}%
]}

\let\figref\relax
\let\tabref\relax
\let\secref\relax
\let\eqnref\relax

\newcommand{\figref}[1]{\hyperref[#1]{Fig.~\ref*{#1}}}
\newcommand{\tabref}[1]{\hyperref[#1]{Table~\ref*{#1}}}
\newcommand{\secref}[1]{\hyperref[#1]{Section~\ref*{#1}}}
\newcommand{\eqnref}[1]{\hyperref[#1]{(\ref*{#1})}}

\begin{document}
\let\WriteBookmarks\relax
\def\floatpagepagefraction{1}
\def\textpagefraction{.001}

\shorttitle{}
\shortauthors{Jin et~al.}

\title[mode=title]{Adaptive Demand-Driven Energy Management of PCM-Integrated District Heating Systems: Operational Flexibility and Techno-Economic Assessment}

\author[1]{Xin Jin}\cormark[1]
\ead{xin.jin1@ntnu.no}
\author[2]{Chunjun Huang}
\author[3]{Pei Huang}
\author[1]{Natasa Nord}

\affiliation[1]{organization={Department of Energy and Process Engineering, Norwegian University of Science and Technology},
    city={Trondheim},
    country={Norway}}
\affiliation[2]{organization={Department of Electrical Sustainable Energy, Delft University of Technology},
    city={Delft},
    country={The Netherlands}}
\affiliation[3]{organization={Department of Engineering Science, Mälardalen University},
    city={Västerås},
    country={Sweden}}

\cortext[cor1]{Corresponding author.}

\begin{abstract}
Latent heat thermal energy storage (LHTES) using phase change materials (PCMs) is a promising solution to shifting heat supply and reducing peak demand in district heating (DH). However, the combined impacts of PCM thermophysical properties and practical control strategies on DH system-level operational and economic performance remain insufficiently understood. To bridge the research gap, this study investigates a PCM-integrated DH system with heat pump assisted waste heat recovery under an adaptive demand-driven (ADD) control strategy to enhance operational flexibility. A dynamic simulation model was developed and the system performance was evaluated against a baseline case and a rule-based control (RBC) approach based on peak-load reduction, operational cost, heat pump performance, and indoor thermal comfort. Furthermore, sensitivity analyses were conducted to examine the influence of PCM thermophysical properties on system performance. The results showed that the RBC can shift peak demand but tends to generate secondary peaks during charging periods. In contrast, the ADD strategy effectively smoothed the heat demand profile and  achieved up to 5.3\% peak load reduction while maintaining thermal comfort. Sensitivity analysis revealed that a phase-change temperature of \SI{80}{\degreeCelsius} and thermal conductivity above \SI{2}{W/(m\cdot K)} achieved a higher peak-load reduction and improved economic performance. Despite the enhanced peak-shaving capability achieved by the proposed control strategy, the system exhibited a payback period of 25.3 years, indicating that further cost reductions and supportive market incentives are required. Nevertheless, the proposed approach provides significant potential for enhancing DH flexibility and supporting the transition toward future low-carbon energy systems.
\end{abstract}

% \begin{highlights}
% \item Adaptive PCM control reduces DH peaks without new charging peaks.
% \item PCM storage maintains indoor comfort and heat-pump operation.
% \item PCM temperature should match the network operating window.
% \item Economic viability depends on peak tariff and storage CAPEX.
% \end{highlights}


\maketitle

% ---------------- Abbreviation ----------------
\nomenclature[A-01-ADD]{\(\rm{ADD}\)}{Adaptive demand driven}
\nomenclature[A-01-BOP]{\(\rm{BOP}\)}{Balance of plant}
\nomenclature[A-02-CAPEX]{\(\rm{CAPEX}\)}{Capital expenditure}
\nomenclature[A-03-DH]{\(\rm{DH}\)}{District heating}
\nomenclature[A-04-DPP]{\(\rm{DPP}\)}{Discounted payback period}
\nomenclature[A-05-HP]{\(\rm{HP}\)}{Heat pump}
\nomenclature[A-06-HTF]{\(\rm{HTF}\)}{Heat transfer fluid}
\nomenclature[A-07-LHTES]{\(\rm{LHTES}\)}{Latent heat thermal energy storage}
\nomenclature[A-08-NPV]{\(\rm{NPV}\)}{Net present value}
\nomenclature[A-09-PCM]{\(\rm{PCM}\)}{Phase change material}
\nomenclature[A-10-SPP]{\(\rm{SPP}\)}{Simple payback period}
\nomenclature[A-11-RBC]{\(\rm{RBC}\)}{Rule based control}
\nomenclature[A-12-TES]{\(\rm{TES}\)}{Thermal energy storage}

% ---------------- Symbols: Latin symbols ----------------
\nomenclature[S-02-A]{\(\rm{A}\)}{Area (m$^{2}$)}
\nomenclature[S-02-C]{\(\rm{C}\)}{Cost (\euro)}
\nomenclature[S-03-cp]{\(\rm{c_p}\)}{Specific heat capacity (J/(kg$\cdot{}^\circ$C))}
\nomenclature[S-03-dh]{\(\rm{d_h}\)}{Hydraulic diameter (m)}
\nomenclature[S-03-E]{\(\rm{E}\)}{Energy (kWh)}
\nomenclature[S-03-f]{\(\rm{f}\)}{Fraction (\%)}
\nomenclature[S-03-Fn]{\(\rm{F_n}\)}{Radiator heat emission coefficient}
\nomenclature[S-03-h]{\(\rm{h}\)}{Heat transfer coefficient (W/(m$^{2}$$\cdot$K))}
\nomenclature[S-05-k]{\(\rm{k}\)}{Thermal conductivity (W/(m$\cdot$K))}
\nomenclature[S-03-L]{\(\rm{L}\)}{Latent heat (kJ/kg)}
\nomenclature[S-04-m]{\(\rm{\dot m}\)}{Mass flow rate (kg/s)}
\nomenclature[S-04-Nu]{\(\rm{Nu}\)}{Nusselt number}
\nomenclature[S-04-Q]{\(\rm{\dot Q}\)}{Heat rate (W)}
\nomenclature[S-06-R]{\(R\)}{Thermal resistance (K/W)}
\nomenclature[S-06-S]{\(S\)}{Saving (\euro)}
\nomenclature[S-06-T]{\(T\)}{Temperature (\(^\circ\mathrm{C}\))}
\nomenclature[S-06-t]{\(t\)}{Time (s)}
\nomenclature[S-04-V]{\(\rm{V}\)}{Volume (m$^3$)}

% ---------------- Symbols: Greek symbols ----------------
\nomenclature[S-26-eta]{\(\eta\)}{Efficiency}
\nomenclature[S-27-rho]{\(\rho\)}{Density (kg/m$^3$)}

% ---------------- Subscripts ----------------
\nomenclature[P-02-ch]{\(\rm{ch}\)}{Charging}
\nomenclature[P-02-dis]{\(\rm{dis}\)}{Discharging}
\nomenclature[P-02-env]{\(\rm{env}\)}{Envelope}
\nomenclature[P-02-ia]{\(\rm{ia}\)}{Indoor air}
\nomenclature[P-04-in]{\(\rm{in}\)}{Inlet}
\nomenclature[P-04-inst]{\(\rm{in}\)}{Installation}
\nomenclature[P-04-l]{\(\rm{l}\)}{Liquid}
\nomenclature[P-04-main]{\(\rm{main}\)}{Maintenance}
\nomenclature[P-04-oa]{\(\rm{oa}\)}{Outdoor air}
\nomenclature[P-04-out]{\(\rm{out}\)}{Outlet}
\nomenclature[P-04-ra]{\(\rm{ra}\)}{Radiator}
\nomenclature[P-04-s]{\(\rm{s}\)}{Solid}
\nomenclature[P-04-th]{\(\rm{th}\)}{Threshold}
\nomenclature[P-04-ven]{\(\rm{ven}\)}{Ventilation}

\section{Introduction}\label{sec:introduction}

District heating (DH) systems play a crucial role in the decarbonization of the building sector by enabling the large-scale utilization of renewable energy and industrial waste heat \cite{Moser2026PeakLoad}. As the share of variable and distributed energy resources continues to increase, DH networks are evolving from conventional heat supply systems toward more flexible and intelligent energy infrastructures \cite{Rahlf2026FlexibleHPDH}. In Nordic countries, where heating demand remains substantial during long winter seasons, DH is regarded as an important pathway for alleviating pressure on the electricity grid \cite{Ding2022LTDHLoadPrediction}. In Norway, DH use has more than doubled over the past decade, with approximately 28.3\% of the generated heat supplied to residential buildings and 52.5\% to service-sector buildings \cite{SSB2025DH}. Nevertheless, the increasing variability of heat demand, particularly during cold-weather periods, often leads to pronounced peak loads that require oversized generation and distribution capacities \cite{Guelpa2017}. Meanwhile, DH tariff structures commonly include demand-based charges associated with peak heat demand, resulting in higher energy costs for end users \cite{Hou2023MPC}. Therefore, reducing and shifting peak loads has become a key strategy for improving the economic performance and operational flexibility of DH networks.
\printNomenclatureTopNextPage

Recently, increasing attention has been paid to integrating thermal energy storage (TES) into DH networks to shift peak load and enhance energy flexibility \cite{Serra2025ExplainableDHForecasting,Ceruti2025DHNStorage}. By storing thermal energy during periods of low demand and releasing it during peak-demand periods, TES can enhance operational flexibility, improve the utilization of renewable and waste heat resources, and lower overall operating costs \cite{DelgadoDiaz2025HybridSTES}. Water-based sensible heat storage has been extensively applied in DH networks due to its simplicity and low cost \cite{Trabert2024}. However, its relatively low energy storage density often results in large storage volumes, which may limit its applicability in space-constrained urban environments and building-level applications \cite{Wan2025}. Compared with sensible heat storage, latent heat thermal energy storage (LHTES) using phase change materials (PCMs) offers significantly higher energy storage density and relatively stable thermal supply. A large amount of thermal energy can be absorbed and released during the phase transition process of PCMs due to their high enthalpy \cite{Agrebi2025HelicalPCM}. Consequently, a smaller storage unit can be used to achieve the same storage capacity, which is particularly advantageous for DH applications where space availability and infrastructure constraints in existing buildings may be limited. In addition, the phase-change process enables thermal energy to be stored and released over a narrow temperature range \cite{Kazaz2026PCS}, providing a more stable heat output for load shifting and flexible operation of DH systems.

Therefore, LHTES has attracted increasing attention as a compact and efficient solution for enhancing the flexibility and operational performance of DH systems. Kollmar et al. developed and experimentally validated a detailed thermo-physical model of a DH and cooling network incorporating a seasonal LHTES system \cite{Kollmar2025_5GDHC}. The ice storage delivered heat during the cold season through the latent heat of water crystallization and provided cooling during summer operation. The model included the thermal interactions among the ice storage, pipes, soil, and building substations. The results showed good agreement between simulations and measurements. However, the study primarily focused on model development and validation and did not investigate storage-property optimization, peak-load reduction performance, or techno-economic impacts of LHTES integration. 
Hlimi et al. developed a numerical model of a PCM-based LHTES system for urban DH applications and investigated the effects of storage design and operating parameters on thermal performance \cite{Hlimi2023_PCM_DH}. The results identified an optimal PCM melting temperature of approximately 70 °C for the studied operating conditions, enabling a stable thermal output of 40 kW for approximately 8 h. However, the study primarily focused on the thermal behaviour and sizing of a standalone PCM storage unit and did not investigate its interaction with DH networks or system-level operation.
Calise et al. proposed a district energy system combining photovoltaics, battery storage, heat pumps (HPs), and PCM thermal storage to convert surplus renewable electricity into thermal or cooling energy for later use \cite{Calise2025_5GDHC_PCM}. Dynamic simulations showed that the system achieved a renewable energy utilization rate of 88.5\%, reduced annual operating costs by more than 84\%, although the high initial investment resulted in a payback period of approximately 19 years. The PCM storage supplied up to 15.8\% of heating demand and 29.5\% of cooling demand. However, the study focused primarily on renewable electricity utilization and power-to-thermal management rather than peak-load reduction and operational flexibility enhancement.
Pans et al. developed a comprehensive modelling framework for fourth-generation DH systems integrated with sensible, latent, and thermochemical TES technologies, enabling system-level evaluation of renewable-based DH networks \cite{Pans2023DH}. The study highlighted the potential of TES to improve the utilization of renewable heat sources and facilitate their integration into DH networks. However, the analysis primarily focused on methodology development and did not investigate the influence of PCM thermophysical properties, storage design parameters, or advanced control strategies on the energy and economic performance of DH systems.

Thomson and Claudio proposed a HP-assisted PCM storage concept for DH network, in which the PCM was charged by a water-source HP and discharged to DH operating at different temperature levels \cite{Thomson2019PCM_DHN}. The results showed that PCM integration could increase HP utilization by 5–8.5\% and improve the low-carbon heat source utilization. Among the investigated materials, sodium acetate trihydrate (SAT) was identified as the most promising PCM due to its favorable thermophysical properties, commercial availability, and cycling stability. However, the study relied on simplified system modelling and did not consider dynamic PCM charging or discharging processes.
Karwacki et al. experimentally investigated a LHTES integrated with a DH system to characterize the dynamic charging and discharging behavior of the storage \cite{Karwacki2025EffectiveEnthalpy}. The results demonstrated that the actual performance of PCM storage depends not only on PCM properties but also on the charging and discharging rates, which significantly affect the utilization of latent heat storage capacity. However, the study primarily focused on the thermal characterization of a single PCM storage unit under controlled experimental conditions. 
Lee et al. experimentally demonstrated a shell-and-tube LHTES system integrated with a DH network in an apartment complex for peak-load shifting \cite{Lee2022_PeakLoad_PCM_DH}. A PCM with a melting temperature of 78 °C was employed to store heat during off-peak periods and release heat during peak-demand periods, achieving a 59.9\% reduction in domestic hot water load fluctuations and a demand response duration of approximately 2.1 h. 
Rinaldi et al. investigated the integration of a decentralized PCM-emulsion TES system into a low-temperature DH network for peak shaving and network enlargement \cite{Rinaldi2024_PCMEmulsion_DH}. The results showed that the proposed storage system could reduce peak demand by up to 38.8\% while enabling additional buildings to be connected to an existing saturated DH network. However, the effects of PCM properties and operational strategies on peak-load reduction performance were not systematically evaluated, and a comprehensive techno-economic assessment was not conducted.

Although LHTES has been widely recognized as a promising solution for enhancing the flexibility of DH systems, several research gaps remain. First, existing studies have primarily focused on storage feasibility, sizing, and energy performance. The influence of key PCM thermophysical properties, such as phase-change temperature, latent heat, and thermal conductivity, on system-level operational and economic performance has not been systematically investigated. Second, most previous studies have evaluated LHTES under fixed operating conditions, while neglecting the interaction between PCM storage and operational control strategies. Third, effective charging and discharging controls are essential for fully exploiting the flexibility potential of TES. Model predictive control can achieve optimal performance in heating systems, but its practical implementation often requires accurate forecasting models, detailed system representations, significant computational resources, and advanced communication infrastructure \cite{Mura2026MPCRBC}, which are challenging to obtain. Alternatively, simpler yet effective control approaches have been developed, which can enhance system flexibility while maintaining low implementation complexity. However, for these alternative control approaches, the combined impacts of PCM thermophysical properties and operational control strategies on peak-load reduction, system flexibility, and techno-economic performance under realistic DH operating conditions remain insufficiently understood. 

To address these research gaps, this study develops a dynamic simulation model of a campus-scale DH system integrated with a shell-and-tube PCM storage tank, a HP utilizing data center waste heat, and a building heating system. The aim of this study is to investigate how PCM thermophysical properties and adaptive operational control influence the energy and economic performance of PCM-integrated DH systems, and to identify suitable PCM characteristics and control strategies for enhancing system flexibility and peak-load reduction. The developed model considers the transient charging and discharging processes of the PCM and the heat transfer interactions among major system components. Using the developed model, the impacts of key PCM thermophysical properties, including phase-change temperature, latent heat, and thermal conductivity on system performance are systematically evaluated. In addition, the novelty of this work lies in the implementation of an adaptive demand-driven (ADD) control strategy to coordinate PCM charging and discharging according to real-time heating demand and DH operating conditions. A further novelty is the integrated analysis of the interaction between PCM thermophysical properties and operational control, revealing their combined influence on peak-load reduction, operational flexibility, and techno-economic performance. The proposed framework provides practical guidance for PCM selection and control strategy development in future DH systems.

The remainder of this paper is organized as follows. Section 2 presents the configuration of the PCM-integrated DH system and the corresponding mathematical modelling approach. Section 3 describes the baseline, rule-based control (RBC), and ADD control strategies and the techno-economic evaluation method. Section 4 presents and discusses the effects of PCM thermophysical properties and operational control on system performance and also provides the techno-economic assessment of the proposed system. Finally, Section 5 summarizes the main conclusions and outlines directions for future work.

\section{System description and modelling}\label{sec:system_model}

\subsection{Investigated district heating system}\label{subsec:case_system}

\begin{figure}[pos=htbp]
    \centering
    \includegraphics[width=0.9\textwidth]{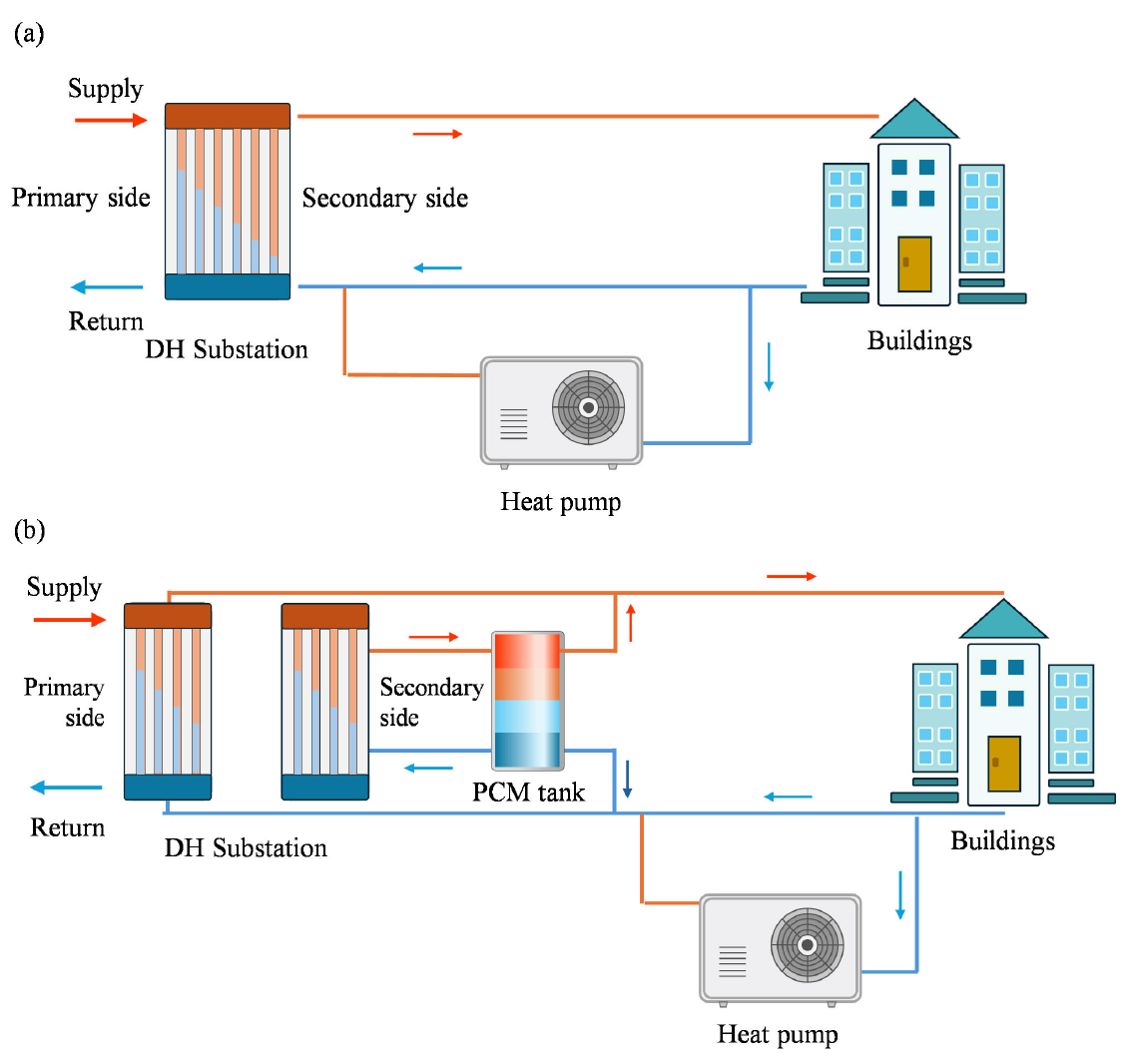}
    \caption{Schematic diagram of the campus district heating system: (a) existing primary district heating system and (b) PCM-integrated district heating system.}
    \label{fig:system_configuration}
\end{figure}

The investigated system is a campus-level two-stage DH network located in Trondheim, Norway (63.43$^\circ$N, 10.40$^\circ$E). As illustrated in \figref{fig:system_configuration}(a), the campus is connected to the city DH network through a main substation, which transfers heat from the primary DH circuit to the secondary campus heating network. Heat is subsequently delivered to campus buildings through building substations. The two-stage configuration enables hydraulic separation between the city DH network and the campus distribution network while providing operational flexibility for local energy management. An HP is integrated into the system to recover low-grade waste heat from the data-center cooling circuit. The recovered thermal energy is upgraded by the HP and supplied to the return side of the main substation through the condenser, thereby reducing the required heat extraction from the city DH network and increasing the utilization of locally available waste heat resources. The primary heating demand considered in this study is space heating supplied through the building radiator systems. The heat demand varies dynamically according to outdoor weather conditions and building occupancy patterns, resulting in pronounced peak-load periods during cold weather. The key component parameters used in the simulations are summarized in \tabref{tab:system_parameters}.

\begin{table}[t]
\caption{Key parameters of the main components in the investigated heating system.}
\label{tab:system_parameters}
\centering
\setlength{\tabcolsep}{4pt}

\begin{tabular*}{\textwidth}{@{\extracolsep{\fill}}
>{\raggedright\arraybackslash}p{\dimexpr0.16\textwidth\relax}
>{\raggedright\arraybackslash}p{\dimexpr0.46\textwidth\relax}
>{\raggedright\arraybackslash}p{\dimexpr0.13\textwidth\relax}
>{\raggedright\arraybackslash}p{\dimexpr0.10\textwidth\relax}
@{}}
\toprule
Component & Parameter & Value & Unit \\
\midrule
Heat pump & Compressor power & 350 & kW \\
 & Condenser-side flow rate & 18 & kg/s \\
 & Evaporator-side flow rate & 36 & kg/s \\
Building & Heated air volume & 1,054,600 & m$^3$ \\
 & Air infiltration rate & 2.45 & h$^{-1}$ \\
 & Heat-recovery temperature efficiency & 58 & \% \\
 & Heat capacity of envelope & $4.5\times10^{10}$ & J/K \\
 & Heat capacity of indoor air & $1.3\times10^{9}$ & J/K \\
 & Heat capacity of internal thermal mass & $2.9\times10^{9}$ & J/K \\
Main substation & Heat transfer area & 420 & m$^2$ \\
 & Capacity & 16.8 & MW \\
\bottomrule
\end{tabular*}
\end{table}

\subsection{PCM storage configuration}\label{subsec:pcm_integration}

To enhance demand-side flexibility and facilitate peak-load shifting, a PCM-based TES system is integrated into the campus DH network, as illustrated in \figref{fig:system_configuration}(b). In the proposed configuration, the main substation consists of two heat exchangers. The first heat exchanger operates conventionally to transfer heat from the city DH network directly to the building substation and satisfy the instantaneous heating demand. The second heat exchanger is used to charge the PCM storage tank during periods of low heat demand or favorable operating conditions. Thermal energy from the DH network is stored in the PCM in a main form of latent heat. During peak-load periods, the PCM storage tank discharges the stored thermal energy to the building substation, thereby supplementing the heating supply and reducing the heat extraction required from the city DH network. This design allows part of the peak demand to be shifted to off-peak periods. 

\begin{figure}[pos=htbp]
    \centering
    \includegraphics[width=0.48\textwidth]{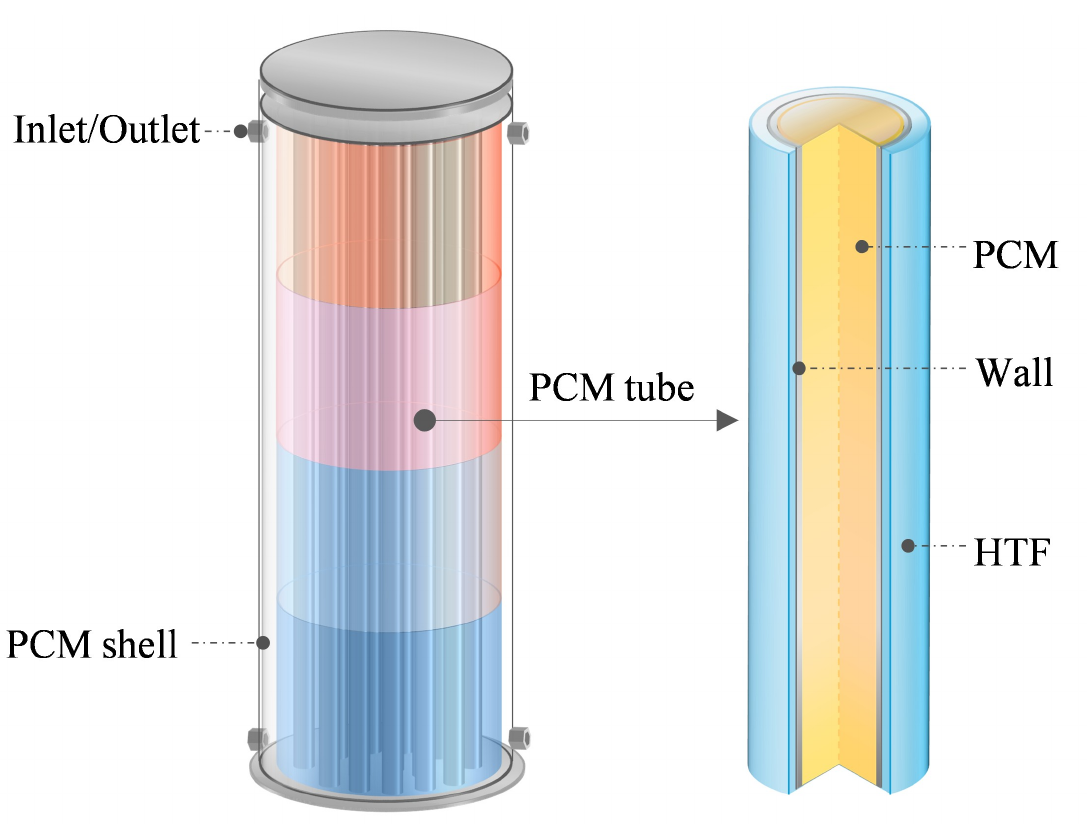}
    \caption{Schematic diagram of the PCM storage tank.}
    \label{fig:PCM_configuration}
\end{figure}

A shell-and-tube PCM TES tank was introduced into the DH system, as seen in \figref{fig:PCM_configuration}. A hybrid sensible and latent TES system was adopted to combine the low cost and high reliability of sensible heat storage with the high energy storage density and nearly isothermal heat transfer characteristics of LHTES. SAT was initially selected as the PCM for the storage tank because of its suitable phase-change characteristics, relatively high energy storage density, commercial availability, and cost-effectiveness \cite{Jayathunga2024}, making it a promising candidate for large-scale DH TES applications. The thermophysical properties of the PCM, heat transfer fluid (HTF), and the corresponding storage tank design parameters adopted in the simulations are summarized in \tabref{tab:pcm_parameters}. The storage capacity was determined from the thermal energy required for peak-load shifting. A target peak-load threshold was defined, and the heat demand exceeding this threshold was expected to be supplied by the PCM storage tank. The required storage capacity was therefore calculated as the cumulative excess thermal energy above the threshold over the considered operation period \eqnref{eq:target_energy}:
\begin{equation}\label{eq:target_energy}
E_{\mathrm{target}} = \int_{t_1}^{t_2} \max\left(\dot{Q}_{\mathrm{load}}(t)-\dot{Q}_{\mathrm{th}},0\right)\,\mathrm{d}t,
\end{equation}
where $E_{\mathrm{target}}$ is the target shifted thermal energy, $\dot{Q}_{\mathrm{load}}$ is the original heating load, and $\dot{Q}_{\mathrm{th}}$ is the target peak-load threshold.

The required PCM volume ($V_{\mathrm{tank}}$) is then estimated using \eqnref{eq:tank_volume}:
\begin{equation}
V_{\mathrm{tank}}
=
\frac{E_{\mathrm{target}}}
{
\eta_u
\left[
\rho_{\mathrm{PCM}}
f_{\mathrm{PCM}}
\left(
c_{p,s}(T_m-T_{\min})
+
L
+
c_{p,l}(T_{\max}-T_m)
\right)
+
\rho_{\mathrm{HTF}}
(1-f_{\mathrm{PCM}})
c_{p,{\mathrm{HTF}}}
(T_{\max}-T_{\min})
\right]
}
\label{eq:tank_volume}
\end{equation}
where $\rho_{\mathrm{PCM}}$ is the PCM density, $\rho_{\mathrm{HTF}}$ is the HTF density, $f_{\mathrm{PCM}}$ is the PCM volume fraction, $c_{p,s}$ and $c_{p,l}$ are the specific heat capacities of solid and liquid PCM, $c_{p,{\mathrm{HTF}}}$ is the specific heat capacity of HTF, $L$ is the latent heat of fusion, $T_m$ is the phase-change temperature, $T_{\min}$ and $T_{\max}$ are the minimum discharging and maximum charging temperatures, and $\eta_{\mathrm{u}}$ is the effective storage utilization factor.

\begin{table}[t]
\caption{Thermophysical properties and design parameters of the PCM storage tank.}
\label{tab:pcm_parameters}
\centering
\setlength{\tabcolsep}{4pt}

\begin{tabular*}{\textwidth}{@{\extracolsep{\fill}}
>{\raggedright\arraybackslash}p{\dimexpr0.22\textwidth\relax}
>{\raggedright\arraybackslash}p{\dimexpr0.38\textwidth\relax}
>{\raggedright\arraybackslash}p{\dimexpr0.16\textwidth\relax}
>{\raggedright\arraybackslash}p{\dimexpr0.12\textwidth\relax}
@{}}
\toprule
Item & Parameter & Value & Unit \\
\midrule
PCM & Phase-change temperature & 60 & $^{\circ}$C \\
 & Latent heat of fusion & 265 & kJ/kg \\
 & Specific heat, solid phase & 2.4 & kJ/(kg$\cdot$K) \\
 & Specific heat, liquid phase & 3.3 & kJ/(kg$\cdot$K) \\
 & Thermal conductivity & 0.25 & W/(m$\cdot$K) \\
 & Density & 1380 & kg/m$^3$ \\
Heat transfer fluid & Specific heat & 4.19 & kJ/(kg$\cdot$K) \\
 & Density & 1000 & kg/m$^3$ \\
 & Thermal conductivity & 0.6 & W/(m$\cdot$K) \\
Encapsulation & Specific heat & 0.5 & kJ/(kg$\cdot$K) \\
 & Density & 7930 & kg/m$^3$ \\
 & Thermal conductivity & 16.3 & W/(m$\cdot$K) \\
Tank & Height & 4.0 & m \\
 & Diameter & 6.5 & m \\
 & PCM volume fraction & 55 & \% \\
\bottomrule
\end{tabular*}
\end{table}

\subsection{Dynamic modelling of the district heating system}\label{subsec:component_models}
The entire modelling framework was developed in Python, where the DH system, PCM storage model, HP, building model, and control strategy were fully integrated into a dynamic simulation environment. 
Weather data for Trondheim were obtained from the PVGIS database developed by the European Commission \cite{PVGISJRC}. In the developed numerical model, the DH energy balance, HP operation, PCM charging and discharging processes, and building heating demand were dynamically evaluated. The developed framework was subsequently used to assess the impacts of control strategies and PCM properties on peak-load shifting performance, operational cost, and overall system flexibility.
\subsubsection{Heat pump performance model}\label{subsubsec:hp_model}

The HP is represented using a semi-empirical performance model fitted from operational data \cite{Nord2021DHProsumer}. The compressor power is expressed as a function of evaporator and condenser inlet and outlet temperatures and flow rates:
\begin{equation}\label{eq:hp_power}
P_{\mathrm{HP}} = aT_{\mathrm{eva,in}} + bT_{\mathrm{eva,out}} + cT_{\mathrm{con,in}} + dT_{\mathrm{con,out}} + e\dot{m}_{\mathrm{eva}} + f\dot{m}_{\mathrm{con}} + g,
\end{equation}
where $P_{\mathrm{HP}}$ is the compressor power, $T_{\mathrm{eva,in}}$ and $T_{\mathrm{eva,out}}$ are the evaporator inlet and outlet temperatures, $T_{\mathrm{con,in}}$ and $T_{\mathrm{con,out}}$ are the condenser inlet and outlet temperatures, and $\dot{m}_{\mathrm{eva}}$ and $\dot{m}_{\mathrm{con}}$ are the evaporator- and condenser-side mass flow rates. The coefficient of performance (COP) is calculated as
\begin{equation}\label{eq:cop}
\mathrm{COP}=\frac{\dot{Q}_{\mathrm{con}}}{P_{\mathrm{HP}}}.
\end{equation}

\begin{figure}[pos=htbp]
    \centering
    \includegraphics[width=0.48\linewidth]{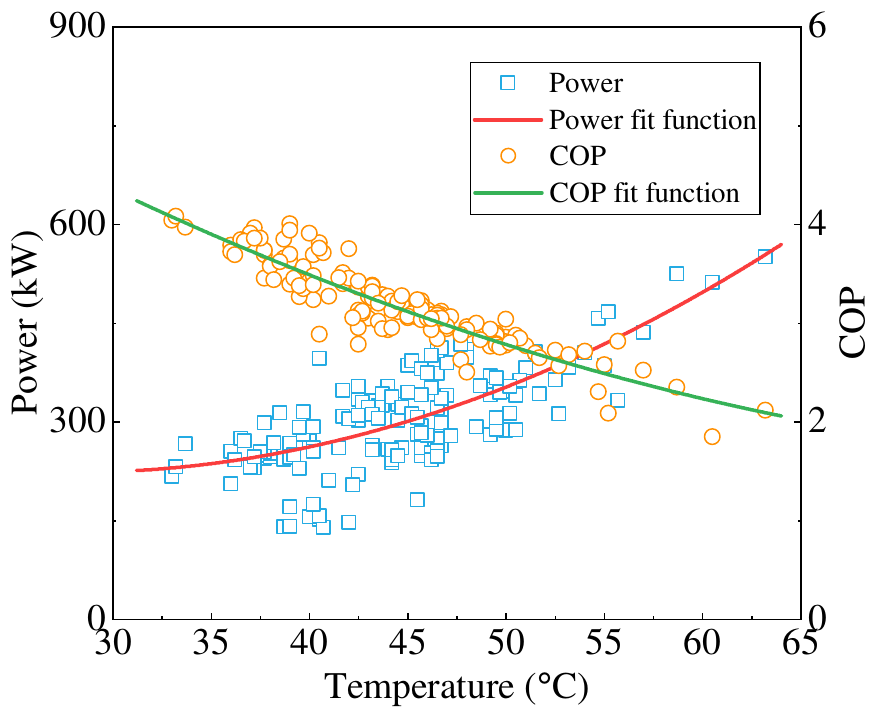}
    \caption{Performance map of the heat pump for data-center heat recovery.}
    \label{fig:hp_performance_map}
\end{figure}

The fitted HP performance map used to relate temperature variation to compressor power and COP is shown in \figref{fig:hp_performance_map}.

\subsubsection{PCM storage model}\label{subsubsec:pcm_model}

The PCM phase-change process is modeled using an effective heat-capacity method. The PCM temperature field is governed by \cite{Tang2026FEA}
\begin{equation}\label{eq:pcm_heat_equation}
\rho_{\mathrm{PCM}} c_{p,\mathrm{eff}}(T)\frac{\partial T_{\mathrm{PCM}}}{\partial t}
= k_{\mathrm{PCM}}\left(\frac{\partial^2 T_{\mathrm{PCM}}}{\partial x^2}+\frac{1}{r}\frac{\partial T_{\mathrm{PCM}}}{\partial r}+\frac{\partial^2 T_{\mathrm{PCM}}}{\partial r^2}\right),
\end{equation}
where $\rho_{\mathrm{PCM}}$ and $k_{\mathrm{PCM}}$ are the PCM density and thermal conductivity, $T_{\mathrm{PCM}}$ is the PCM temperature, $t$ is time (s), $x$ is the axial coordinate, and $r$ is the radial coordinate.

The effective heat capacity ($c_{p,\mathrm{eff}}(T)$) is defined as
\begin{equation}\label{eq:cp_eff}
c_{p,\mathrm{eff}}(T) =
\begin{cases}
 c_{p,s}, & T<T_s, \\
 c_{p,s}+\dfrac{L}{T_l-T_s}, & T_s\leq T\leq T_l, \\
 c_{p,l}, & T>T_l,
\end{cases}
\end{equation}
where $T_s$ and $T_l$ are the lower and upper bounds of the phase-change temperature interval.

The HTF temperature is described by
\begin{equation}\label{eq:htf_model}
\rho_{\mathrm{HTF}}c_{p,\mathrm{HTF}}\left(\frac{\partial T_{\mathrm{HTF}}}{\partial t}+v\frac{\partial T_{\mathrm{HTF}}}{\partial x}\right)
= k_{\mathrm{HTF}}\frac{\partial^2 T_{\mathrm{HTF}}}{\partial x^2}
- h_{\mathrm{HTF}}A_{\mathrm{conv}}\left(T_{\mathrm{HTF}}-T_{\mathrm{wall}}\right),
\end{equation}
with
\begin{equation}\label{eq:heat_transfer_coeff}
h_{\mathrm{HTF}}=\frac{Nu\,k_{\mathrm{HTF}}}{d_h}.
\end{equation}
where $A_{\mathrm{conv}}$ is the convective heat transfer area between the heat HTF and the tube wall, $Nu$ is the Nusselt number, and $d_h$ is the hydraulic diameter of the tube.

The temperature of the tube wall is governed by:
\begin{equation}\label{eq:wall_model}
\rho_{\mathrm{wall}} c_{p,\mathrm{wall}}
\frac{\partial T_{\mathrm{wall}}}{\partial t}
=
k_{\mathrm{wall}}
\left(
\frac{\partial^2 T_{\mathrm{wall}}}{\partial r^2}
+
\frac{1}{r}
\frac{\partial T_{\mathrm{wall}}}{\partial r}
\right)
+
S_{\mathrm{wall}}
\end{equation}
where $\rho_{\mathrm{wall}}$ is the density of the tube wall material,
$c_{p,\mathrm{wall}}$ is the specific heat capacity of the tube wall material,
$T_{\mathrm{wall}}$ is the tube wall temperature, $k_{\mathrm{wall}}$ is the thermal conductivity of the tube wall material, and $S_{\mathrm{wall}}$ is the source term representing the convective heat transfer between the HTF and the tube wall.

The heat rate charged into or discharged from the PCM storage tank is evaluated as
\begin{equation}\label{eq:pcm_energy}
\dot Q_{\mathrm{PCM}}=\dot{m}_{\mathrm{HTF}}c_{p,\mathrm{HTF}}\left(T_{\mathrm{out}}-T_{\mathrm{in}}\right)
\end{equation}
where $T_{\mathrm{in}}$ and $T_{\mathrm{out}}$ are the inlet and outlet temperatures of the PCM storage tank

\subsubsection{Building thermal model}\label{subsubsec:building_model}
The case study is conducted on the DH system of the NTNU campus, where a representative building is considered as the campus heating demand. The building heating load is calculated dynamically from the lumped-parameter resistance--capacitance thermal network based on the indoor thermal balance. The envelope, indoor air, and internal thermal mass temperatures are obtained from
\begin{equation}
C_{\mathrm{env}}\frac{\mathrm{d}T_{\mathrm{env}}}{\mathrm{d}t}
=
\frac{T_{\mathrm{oa}}-T_{\mathrm{env}}}{R_{o,e}}
+\frac{T_{\mathrm{in}}-T_{\mathrm{env}}}{R_{i,e}}
\label{eq:env_temp}
\end{equation}

\begin{equation}
C_{\mathrm{ia}}\frac{\mathrm{d}T_{\mathrm{ia}}}{\mathrm{d}t}
=
\frac{T_{\mathrm{env}}-T_{\mathrm{ia}}}{R_{i,e}}
+\frac{T_{\mathrm{oa}}-T_{\mathrm{ia}}}{R_{\mathrm{win}}}
+\frac{T_{\mathrm{mass}}-T_{\mathrm{ia}}}{R_{\mathrm{mass}}}
+\frac{T_{\mathrm{oa}}-T_{\mathrm{ia}}}{R_{\mathrm{ven}}}
+\dot{Q}_{\mathrm{rad}}
+\dot{Q}_{\mathrm{ven}}
+\dot{Q}_{\mathrm{int}}
\label{eq:indoor_temp}
\end{equation}

\begin{equation}
C_{\mathrm{mass}}\frac{\mathrm{d}T_{\mathrm{mass}}}{\mathrm{d}t}
=
\frac{T_{\mathrm{ia}}-T_{\mathrm{mass}}}{R_{\mathrm{mass}}}
\label{eq:mass_temp}
\end{equation}
where $T_{\mathrm{env}}$, $T_{\mathrm{ia}}$, $T_{\mathrm{mass}}$, and $T_{\mathrm{oa}}$ are the envelope, indoor air, internal thermal mass, and outdoor air temperatures. The terms $\dot{Q}_{\mathrm{rad}}$, $\dot{Q}_{\mathrm{ven}}$, and $\dot{Q}_{\mathrm{int}}$ represent heat supplied by radiators, ventilation heat flow, and internal heat gains.

The ventilation heat loss ($\dot{Q}_{\mathrm{ven}}$) is calculated as:
\begin{equation}
\dot{Q}_{\mathrm{ven}} =
0.33 \dot{V}_{\mathrm{air}}
(1-\eta_T)
\left(T_{\mathrm{ia,ref}}-T_{\mathrm{oa}}\right)
\end{equation}
where $\dot{V}_{\mathrm{air}}$ is the ventilation airflow rate, 
$\eta_T$ is the heat recovery efficiency of the ventilation system, 
$T_{\mathrm{ia,ref}}$ is the indoor air reference temperature.

The total heat output from the radiator system ($\dot{Q}_{\mathrm{ra}}$) is expressed as:

\begin{equation}
\dot{Q}_{\mathrm{ra}} =
\sum_{n=1}^{N} \dot{Q}_n
\end{equation}
where $\dot{Q}_n$ is the heat output from radiator segment $n$, 
and $N$ is the total number of radiator segments. 

The heat output from radiator segment $n$ is calculated by:
\begin{equation}
\dot{Q}_n =
K_n F_n
\left(T_n-T_{\mathrm{ia}}\right)^{a+1}
\end{equation}
where $K_n$ is the heat transfer coefficient of radiator segment $n$, 
$F_n$ is the heat transfer area of segment $n$, 
$T_n$ is the water temperature of radiator segment $n$, 
and $a$ is the empirical radiator heat emission exponent.

The dynamic temperature variation of each radiator segment is described as:
\begin{equation}
C_{\mathrm{ra}}
\frac{dT_n}{dt}
=
c_{p,HTF}\dot{m}_{\mathrm{ra}}
\left(T_{n-1}-T_n\right)
-
\dot{Q}_n
\end{equation}
where $C_{\mathrm{ra}}$ is the thermal capacitance of the radiator segment, 
$\dot{m}_{\mathrm{ra}}$ is the radiator water mass flow rate, 
and $T_{n-1}$ is the water temperature of the upstream radiator segment.

\subsubsection{District heating system energy balance}

The heat supplied from the city DH network through the main substation was calculated as:
\begin{equation}
\dot Q_{\mathrm{DH}} =
\dot m_{\mathrm{DH}} c_{p,HTF}
\left(
T_{\mathrm{sup,DH}} - T_{\mathrm{ret,DH}}
\right)
\label{eq:dh_heat}
\end{equation}
where $\dot Q_{\mathrm{DH}}$ is the heat rate extracted from the city DH network, $\dot m_{\mathrm{DH}}$ is the mass flow rate of the DH HTF, and $T_{\mathrm{sup,DH}}$ and $T_{\mathrm{ret,DH}}$ are the supply and return temperatures of the DH circuit, respectively.

For the baseline system without PCM storage, the heat supplied from the DH network and the HP should satisfy the building heating demand:
\begin{equation}
\dot Q_{\mathrm{DH,base}} + \dot Q_{\mathrm{HP}}
=
\dot Q_{\mathrm{load}}
\label{eq:baseline_balance}
\end{equation}
where $\dot Q_{\mathrm{load}}$ is the building heating demand and $\dot Q_{\mathrm{HP}}$ is the heat recovered by the HP.

When the PCM storage tank is integrated, the system energy balance becomes:
\begin{equation}
\dot Q_{\mathrm{DH}}
+
\dot Q_{\mathrm{HP}}
+
\dot Q_{\mathrm{PCM,dis}}
=
\dot Q_{\mathrm{load}}
+
\dot Q_{\mathrm{PCM,ch}}
\label{eq:pcm_balance}
\end{equation}
where $\dot Q_{\mathrm{PCM,ch}}$ and $\dot Q_{\mathrm{PCM,dis}}$ are the charging and discharging heat rates of the PCM storage tank, respectively. 

To reduce the influence of short-term operational fluctuations, the peak demand is defined as the average value of the three highest load peaks observed during the evaluation period. The peak heat load imposed on the city DH network was determined as:
\begin{equation}
\dot Q_{\mathrm{peak}} =
\frac{\dot Q_{\mathrm{DH}}^{(1)}+\dot Q_{\mathrm{DH}}^{(2)}+\dot Q_{\mathrm{DH}}^{(3)}}{3}
\label{eq:peak_load}
\end{equation}
where $Q_{\mathrm{DH}}^{(1)}$, $Q_{\mathrm{DH}}^{(2)}$, and $Q_{\mathrm{DH}}^{(3)}$ are the three highest DH load peaks during the evaluation period.

The peak-load shifting rate was then calculated by comparing the peak DH demand before and after integrating the PCM storage system:
\begin{equation}
\eta_{\mathrm{shift}} =
\frac{
\dot Q_{\mathrm{peak,base}}
-
\dot Q_{\mathrm{peak,PCM}}
}{
\dot Q_{\mathrm{peak,base}}
}
\times 100\%
\label{eq:peak_shifting_rate}
\end{equation}
where $\dot Q_{\mathrm{peak,base}}$ and $\dot Q_{\mathrm{peak,PCM}}$ are the peak heat powers extracted from the city DH network in the baseline and PCM-integrated cases, respectively.

\section{Control strategy and evaluation method}\label{sec:control_strategy}

The PCM integrated DH system is evaluated under different control strategies to investigate its capability for improving operational flexibility and reducing peak demand. The system performance is assessed in terms of peak-load shifting effectiveness, total operating cost, indoor thermal comfort, HP performance, and overall system flexibility.

\subsection{Baseline control}\label{subsec:baseline_control}

\begin{figure}[pos=!htbp]
    \centering
    \includegraphics[width=0.92\textwidth]{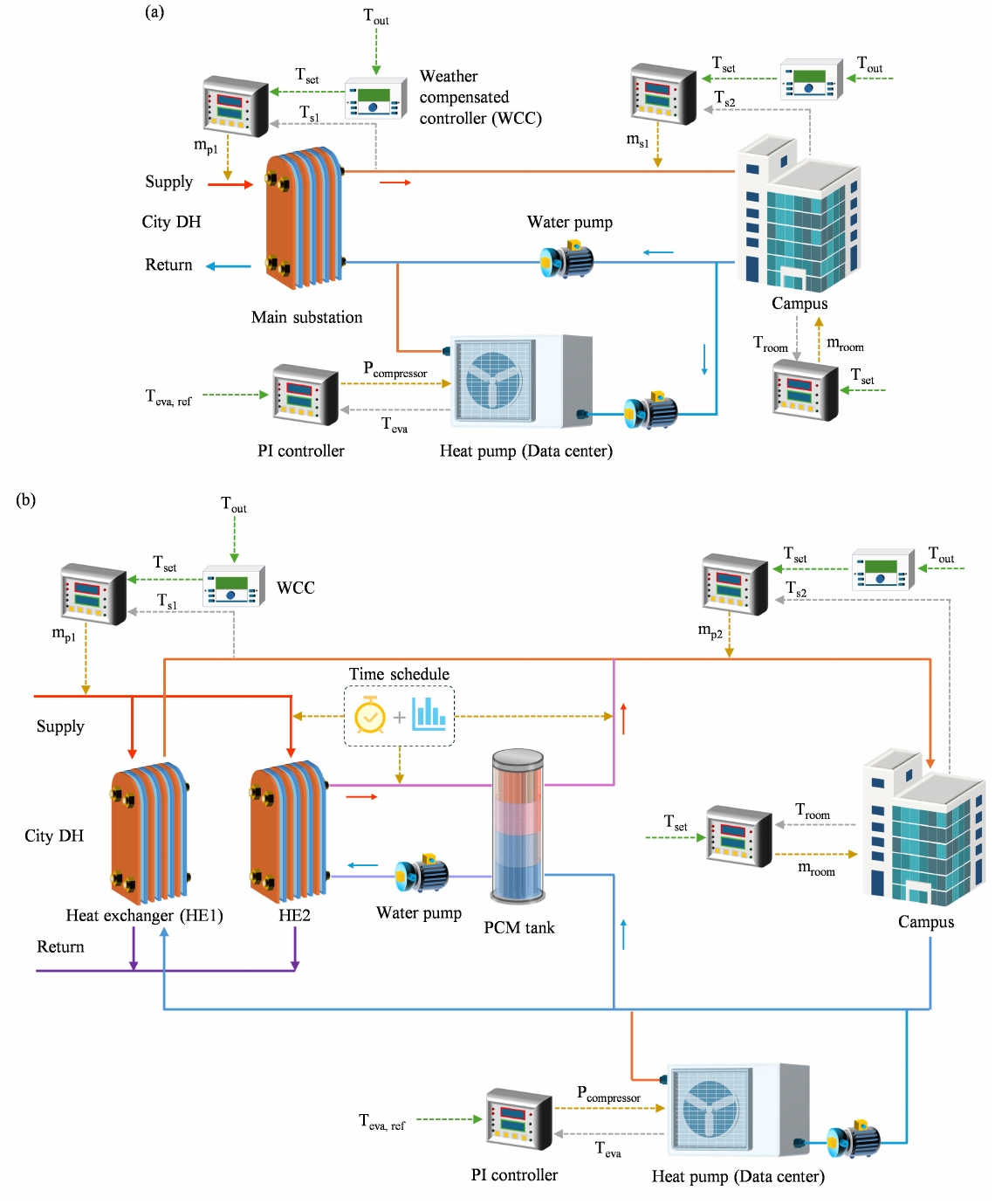}
    \caption{Schematic diagram of the control schemes: (a) baseline control of the existing district heating system and (b) schedule-based rule-based thermal storage control.}
    \label{fig:control_schemes}
\end{figure}

As shown in \figref{fig:control_schemes}(a), the baseline system uses weather-compensated control to determine the secondary-side supply temperature setpoint according to the outdoor temperature. A proportional--integral (PI) controller regulates the primary-side flow rate to track the desired supply temperature. At the building substation, another weather-compensated controller determines the secondary supply temperature setpoint, and a PI controller adjusts the primary-side flow rate. Indoor temperature is maintained by modulating radiator group flow rates through an additional PI controller. The HP controller maintains the evaporator outlet temperature close to \SI{7}{\degreeCelsius} by adjusting the compressor power. Anti-windup and deadband filtering are included to avoid frequent hunting.

\subsection{Schedule-based rule-based PCM control}\label{subsec:rbc_control}

The schedule-based RBC, illustrated in \figref{fig:control_schemes}(b), is used as a benchmark for PCM operation. In addition to the existing control strategy of the DH system, a predefined time schedule is implemented to govern the charging and discharging operation of the PCM storage tank. Specifically, it is charged during off-peak hours from 20:00 to 05:00 the following day at a designed power rate of 3 MW. Subsequently, it is discharged during peak hours from 05:00 to 20:00 at the same power rate, thereby shaving the peak load for the DH. This charge and discharge cycle of PCM storage tank was intended to reduce the reliance on the primary heat source during periods of high demand.

\subsection{Adaptive demand-driven PCM control}\label{subsec:add_control}

\begin{figure*}[pos=htbp]
    \centering
    \includegraphics[width=0.95\textwidth]{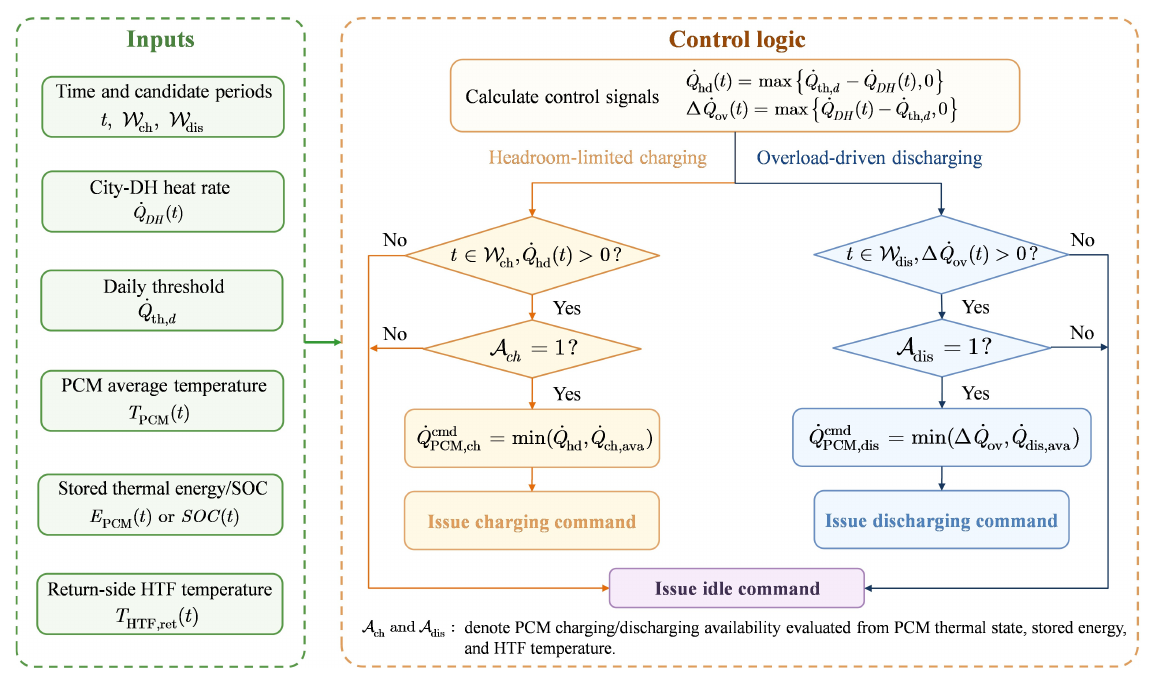}
    \caption{Flowchart of the adaptive demand-driven PCM thermal storage control strategy.}
    \label{fig:add_flowchart}
\end{figure*}

The proposed ADD control strategy operates the PCM storage tank through three modes: charging, discharging, and idle, as illustrated in \figref{fig:add_flowchart}. Unlike the schedule-based RBC strategy, which charges and discharges the storage according to fixed time windows and predefined power rates, the ADD controller updates the PCM operating mode according to the instantaneous city DH heat rate, the daily peak-load threshold, and the thermal state of the PCM storage. In \figref{fig:add_flowchart}, the left branch corresponds to headroom-limited charging during low-load periods, while the right branch corresponds to overload-driven discharging during peak-load periods. If neither branch satisfies the operating conditions, the PCM tank remains idle. The main control settings are summarized in \tabref{tab:add_parameters}.

For each day $d$, a daily peak-load threshold is defined as
\begin{equation}\label{eq:add_daily_threshold}
\dot Q_{\mathrm{th},d}
=
(1-\alpha_{\mathrm{sh}})\dot Q_{\mathrm{daily,max},d},
\end{equation}
where $\alpha_{\mathrm{sh}}$ is the target peak-load reduction ratio and $\dot Q_{\mathrm{daily,max},d}$ is the reference daily peak heat rate. This threshold is used as the common reference for both charging and discharging decisions.

During the candidate charging period, the PCM storage tank is allowed to charge only when the city DH heat rate is below the daily threshold. The available charging headroom is calculated as
\begin{equation}\label{eq:add_charging_headroom}
\dot Q_{\mathrm{hd}}(t)
=
\max\left(\dot Q_{\mathrm{th},d}-\dot Q_{\mathrm{DH}}(t),0\right),
\end{equation}
where $\dot Q_{\mathrm{DH}}(t)$ is the instantaneous city DH heat rate used for the ADD control decision. This headroom-limited charging logic prevents the PCM storage tank from creating additional off-peak demand peaks.

During the candidate discharging period, the controller evaluates the overload level above the daily threshold:
\begin{equation}\label{eq:add_overload_signal}
\Delta \dot Q_{\mathrm{ov}}(t)
=
\max\left(\dot Q_{\mathrm{DH}}(t)-\dot Q_{\mathrm{th},d},0\right).
\end{equation}
This overload signal is used to activate PCM discharging only when the city DH demand approaches or exceeds the daily threshold. Therefore, the PCM storage tank releases heat according to the actual peak-shaving need rather than discharging at a fixed rate throughout the entire daytime period.

The operating mode is selected as
\begin{equation}\label{eq:add_mode_selection}
M(t)=
\begin{cases}
\mathrm{CH}, &
t\in\mathcal{W}_{\mathrm{ch}},\
\dot Q_{\mathrm{hd}}(t)>0,\
\mathcal{A}_{\mathrm{ch}}(t)=1,\\
\mathrm{DIS}, &
t\in\mathcal{W}_{\mathrm{dis}},\
\Delta \dot Q_{\mathrm{ov}}(t)>0,\
\mathcal{A}_{\mathrm{dis}}(t)=1,\\
\mathrm{IDLE}, & \mathrm{otherwise},
\end{cases}
\end{equation}
where $\mathcal{W}_{\mathrm{ch}}$ and $\mathcal{W}_{\mathrm{dis}}$ are the candidate charging and discharging periods, respectively. The binary terms $\mathcal{A}_{\mathrm{ch}}(t)$ and $\mathcal{A}_{\mathrm{dis}}(t)$ denote the charging and discharging availability of the PCM storage tank. Specifically, $\mathcal{A}_{\mathrm{ch}}(t)=1$ indicates that the storage can still accept heat, while $\mathcal{A}_{\mathrm{dis}}(t)=1$ indicates that sufficient stored energy and thermal driving potential are available for effective discharge.

When the controller enters the charging mode, the charging command is limited by both the available network headroom and the instantaneous charging capability of the PCM tank:
\begin{equation}\label{eq:add_charge_command}
\dot Q_{\mathrm{PCM,ch}}^{\mathrm{cmd}}(t)
=
\min\left[
\dot Q_{\mathrm{hd}}(t),
\dot Q_{\mathrm{ch,ava}}(t)
\right],
\end{equation}
where $\dot Q_{\mathrm{ch,ava}}(t)$ denotes the available charging capability of the PCM tank. It is evaluated from the dynamic PCM storage model according to the current PCM temperature, stored thermal energy, and the charging temperature target.

When the controller enters the discharging mode, the discharge command is determined by the overload signal and limited by the available discharge capability:
\begin{equation}\label{eq:add_discharge_command}
\dot Q_{\mathrm{PCM,dis}}^{\mathrm{cmd}}(t)
=
\min\left[
\Delta \dot Q_{\mathrm{ov}}(t),
\dot Q_{\mathrm{dis,ava}}(t)
\right],
\end{equation}
where $\dot Q_{\mathrm{dis,ava}}(t)$ denotes the available discharging capability of the PCM storage tank. It is evaluated according to the stored thermal energy and the thermal driving potential between the PCM and the return-side HTF. Therefore, the return-side HTF temperature is used to assess whether effective PCM-to-HTF heat transfer can be achieved during discharging.

The actual heat exchanged by the PCM storage tank is then calculated by the dynamic PCM storage model introduced in \secref{subsubsec:pcm_model}. Through this mode-selection and command-limiting logic, the ADD strategy shifts part of the city DH heat extraction from peak-load periods to low-load periods while avoiding charging-induced peaks. This makes the proposed controller different from the schedule-based RBC strategy: the candidate time windows are retained for practical operation, but the actual charging and discharging actions are governed by network headroom, overload level, and PCM storage availability.

% \begin{table}[h]
% \centering
% \caption{Key control parameters adopted in the PCM-assisted ADD control strategy.}
% \begin{tabular}{p{7cm}p{7cm}}
% \hline
% Parameter & Value / Description \\
% \hline
% Peak-load threshold &
% $P_{\mathrm{grid,limit}}
% =
% (1-0.15)\times P_{\mathrm{daily,max}}$ \\
% Discharge activation threshold &
% $0.985 \times P_{\mathrm{grid,limit}}$ \\
% Discharge deactivation threshold &
% $0.94 \times P_{\mathrm{grid,limit}}$ \\
% Hysteresis band &
% 4.5\% of the daily peak-load threshold \\
% % Maximum charging power &
% % 6 MW \\
% % Maximum discharging power &
% % 14 MW \\
% Minimum discharge temperature difference &
% 0.8$^\circ$C \\
% Maintaining discharge temperature difference &
% 0.2$^\circ$C \\
% Charging period &
% 20:00--05:00 \\
% Discharging period &
% 05:00--20:00 \\
% Simulation time step &
% 10 s \\
% Control update interval &
% 10 s \\
% \hline
% \end{tabular}
% \label{tab:add_parameters}
% \end{table}

\begin{table}[h]
\centering
\caption{Key numerical settings adopted in the control strategy for PCM integrated DH system.}
\begin{tabular*}{\linewidth}{@{\extracolsep{\fill}}p{0.48\linewidth}p{0.47\linewidth}@{}}
\hline
Parameter & Value \\
\hline
Peak-load reduction target, $\alpha_{\mathrm{sh}}$ &
15\% \\

Threshold update interval &
1 day \\

Candidate charging period, $\mathcal{W}_{\mathrm{ch}}$ &
20:00--05:00 \\

Candidate discharging period, $\mathcal{W}_{\mathrm{dis}}$ &
05:00--20:00 \\

Maximum charging power &
2 MW \\

Maximum discharging power &
2 MW \\

PCM charging temperature target &
95$^\circ$C \\

Simulation interval &
10 s \\
\hline
\end{tabular*}
\label{tab:add_parameters}
\end{table}

% \begin{table}[h]
% \centering
% \caption{Key control settings adopted in the ADD control strategy.}
% \begin{tabular*}{0.8\linewidth}{@{\extracolsep{\fill}}p{0.57\linewidth}p{0.38\linewidth}@{}}
% \hline
% Parameter & Value / Definition \\
% \hline
% Peak-load reduction target, $\alpha$ &
% 15\% \\

% PCM discharge temperature limit &
% 45$^\circ$C \\

% %Daily peak-load threshold, $\dot Q_{\mathrm{th},d}$ &
% %$(1-\alpha_{\mathrm{sh}})\dot Q_{\mathrm{daily,max},d}$ \\

% %Charging admissible condition &
% %$\dot Q_{\mathrm{DH}}(t)<\dot Q_{\mathrm{th},d}$ \\

% %Charging headroom, $\dot Q_{\mathrm{hd}}(t)$ &
% %$\dot Q_{\mathrm{th},d}-\dot Q_{\mathrm{DH}}(t)$ \\

% %Discharging demand signal, $\Delta \dot Q_{\mathrm{ov}}(t)$ &
% %$\max\left(\dot Q_{\mathrm{DH}}(t)-\dot Q_{\mathrm{th},d},0\right)$ \\

% PCM charging temperature target &
% 90$^\circ$C \\

% Minimum discharge temperature difference &
% 1.0$^\circ$C \\

% Maximum charging power &
% 5 MW \\

% Maximum discharging power &
% 5 MW \\

% Candidate charging period &
% 20:00--05:00 \\

% Candidate discharging period &
% 05:00--20:00 \\

% Simulation and control update interval &
% 10 s \\
% \hline
% \end{tabular*}

% %\vspace{1mm}
% %\begin{minipage}{\linewidth}
% %\footnotesize
% %\textit{Note:} $\dot Q_{\mathrm{DH}}(t)$ denotes the instantaneous heat rate extracted from the city DH network used for the ADD control decision; subscript $d$ denotes the corresponding day.
% %\end{minipage}
% \label{tab:add_parameters}
% \end{table}

\subsection{Techno-economic evaluation method}\label{subsec:techno_economic_method}
The DH operating cost is calculated by combining the total energy cost and the peak-demand cost:
\begin{equation}\label{eq:operating_cost}
C_{\mathrm{op}} = C_{\mathrm{energy}}+C_{\mathrm{peak}}
=\sum_{t}p_e\dot{Q}_{\mathrm{DH}}(t)\Delta t
+p_{\mathrm{peak}}\dot{Q}_{\mathrm{peak}}
\end{equation}
where $C_{\mathrm{op}}$ is the total operating cost over the evaluation period, $C_{\mathrm{energy}}$ is the energy cost, $C_{\mathrm{peak}}$ is the peak-demand cost, $p_e$ is the heat energy price of 0.060~EUR/kWh, $p_{\mathrm{peak}}$ is the peak-demand tariff of 3.25~EUR/kW$\cdot$month \cite{Du2025}, $\Delta t$ is the simulation time step.

The annual operating-cost saving ($S_{\mathrm{ann}}$) is calculated as
\begin{equation}\label{eq:annual_saving}
S_{\mathrm{ann}}=
C_{\mathrm{op,base}}^{\mathrm{ann}}
-
C_{\mathrm{op,PCM}}^{\mathrm{ann}},
\end{equation}
where $C_{\mathrm{op,base}}^{\mathrm{ann}}$ is the annual operating cost of the baseline DH system, and $C_{\mathrm{op,PCM}}^{\mathrm{ann}}$ is the annual operating cost of the PCM-integrated DH system under the selected control strategy.

The capital expenditure (CAPEX) of the PCM storage subsystem is estimated using a component-level cost breakdown, as expressed in Eq.~(\ref{eq:capex_total}), and the economic parameters for CAPEX and economic sensitivity analysis is listed in Table~\ref{tab:economic_parameters}.

\begin{equation}\label{eq:capex_total}
\mathrm{CAPEX} =
C_{\mathrm{PCM}}
+
C_{\mathrm{tank}}
+
C_{\mathrm{pump,BOP}}
+
C_{\mathrm{inst}}
\end{equation}
where $C_{\mathrm{PCM}}$ is the PCM material cost, $C_{\mathrm{tank}}$ is the tank and insulation cost, $C_{\mathrm{pump,BOP}}$ is the cost allowance for pumps, valves, local piping and balance-of-plant components, $C_{\mathrm{inst}}$ is the delivery and installation cost.

The tank and insulation cost are estimated as fractions of the PCM material cost: 
\begin{equation}\label{eq:capex_tank}
C_{\mathrm{tank}} = f_{\mathrm{tank}} C_{\mathrm{PCM}}
\end{equation}
where $f_{\mathrm{tank}}$ is the tank and insulation cost factor.

The balance-of-plant (BOP) including pump, valve, and local piping cost are estimated as:
\begin{equation}\label{eq:capex_bop}
C_{\mathrm{pump,BOP}}
=
f_{\mathrm{BOP}}
\left(
C_{\mathrm{PCM}}
+
C_{\mathrm{tank}}
\right)
\end{equation}
where $f_{\mathrm{BOP}}$ is the auxiliary pump and balance-of-plant cost factor.

Delivery, installation, and control integration are then estimated as:
\begin{equation}\label{eq:capex_inst}
C_{\mathrm{inst}}
=
f_{\mathrm{inst}}
\left(
C_{\mathrm{PCM}}
+
C_{\mathrm{tank}}
+
C_{\mathrm{pump,BOP}}
\right)
\end{equation}

The annual maintenance cost is estimated as a fraction of the initial investment cost. The annual net cash flow is then calculated as
\begin{equation}\label{eq:net_saving}
S_{\mathrm{net}}
=
S_{\mathrm{ann}}
-
C_{\mathrm{main}},
\end{equation}
where $S_{\mathrm{net}}$ is the annual net cash flow after deducting the annual operation and maintenance cost, $S_{\mathrm{ann}}$ is the annual operating-cost saving, and $C_{\mathrm{main}}$ is the annual maintenance cost.

The simple payback period (SPP) is determined as:
\begin{equation}
\mathrm{SPP}=\frac{\mathrm{CAPEX}}
{S_{\mathrm{net}}}
\end{equation}

The corresponding net present value (NPV) over the project lifetime is calculated as:
\begin{equation}
\mathrm{NPV}
=
-\mathrm{CAPEX}
+
\sum_{t=1}^{N}
S_{\mathrm{net}}
\end{equation}
where $N$ is the project lifetime

%The discounted payback period (DPP) is determined as the first year in which the cumulative discounted cash flow becomes non-negative:
%\begin{equation}\label{eq:dpp}
%\mathrm{DPP}
%=
%\min
%\left\{
%n:
%-C_0
%+
%\sum_{y=1}^{n}
%\frac{
%S_{\mathrm{net}}
%}{
%(1+r)^y
%}
%\geq 0
%\right\},
%\end{equation}

%The net present value (NPV) over the project lifetime is calculated as
%\begin{equation}\label{eq:npv}
%\mathrm{NPV}
%=
%-C_0
%+
%\sum_{y=1}^{N}
%\frac{
%S_{\mathrm{net}}
%}{
%(1+r)^y
%},
%\end{equation}
%where $N$ is the project lifetime, $y$ is the year index, and $r$ is the discount rate. 

\begin{table}[pos=htbp]
\caption{Economic parameters used for the component-level CAPEX and sensitivity analysis.}
\label{tab:economic_parameters}
\centering
\setlength{\tabcolsep}{4pt}
\begin{tabular*}{0.75\textwidth}{@{\extracolsep{\fill}}
>{\raggedright\arraybackslash}p{0.3\textwidth}
>{\raggedright\arraybackslash}p{0.25\textwidth}
>{\raggedright\arraybackslash}p{0.1\textwidth}
@{}}
\toprule
Item & Reference value & Basis \\
\midrule
PCM material, $c_{\mathrm{PCM}}$
& 1.0 €/kg
& \citep{Xu2025,Mekrisuh2025} \\

Tank and insulation, $f_{\mathrm{tank}}$
& 30\%
& \citep{Selvakumar2025LHTES} \\

Pump and local BOP factor, $f_{\mathrm{BOP}}$
& 15\%
& \\

Installation factor, $f_{\mathrm{inst}}$
& 20\%
& \citep{Selvakumar2025LHTES} \\

Annual Maintenance cost, $f_{\mathrm{main}}$
& 2\%
& \citep{Jin2021} \\

Project lifetime, $N$
& 30 years
& \\

%Discount rate, $r$
%& 4\%
%& \citep{EC2021,DEA2024} \\
\bottomrule
\end{tabular*}
\end{table}

\section{Results}\label{sec:results}
\subsection{Comparison results of baseline, RBC, and ADD Performance}
\subsubsection{Indoor thermal condition and heat-pump operation}\label{subsec:comfort_hp}

\begin{figure}[pos=htbp]
    \centering
    \includegraphics[width=0.8\linewidth]{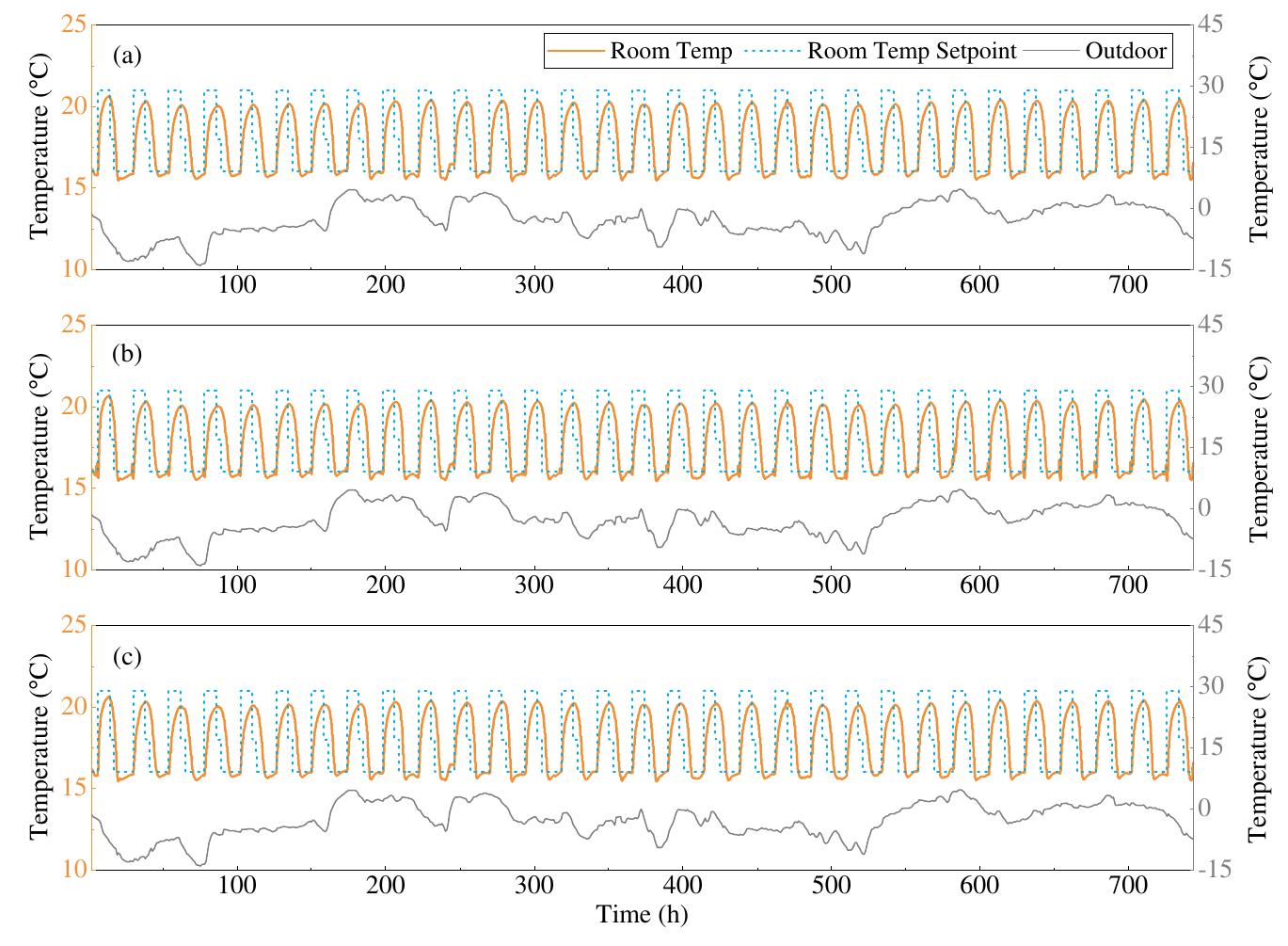}
    \caption{Indoor air temperature compared against scheduled setpoint for (a) the baseline system, (b) the PCM-integrated district heating system under rule-based control, and (c) the PCM-integrated district heating system under adaptive demand-driven control.}
    \label{fig:indoor_temperature}
\end{figure}

The system is first simulated over a continuous and representative 31-day winter period to evaluate and compare the performance of different simulation cases. The same weather conditions and heating demand profile are applied to all cases to ensure a consistent basis for comparison. Before comparing the peak-shaving and cost performance of the different control strategies, the indoor thermal condition and HP operation are checked to ensure that the cases are evaluated under comparable thermal comfort reliability. \figref{fig:indoor_temperature} shows the temperature evolution of indoor air temperature compared against the scheduled setpoint over the simulation period. The indoor room temperature shows a similar trend across all cases, indicating consistent thermal behavior of the system under different configurations. In all scenarios, the indoor temperature is able to flexibly track the setpoint and remains within a suitable thermal comfort range of 19--21~$^{\circ}$C during occupied hours in the campus. During unoccupied periods, the room temperature is effectively reduced, contributing to the reduction of unnecessary energy use. 

\begin{figure}[pos=!htbp]
    \centering
    \includegraphics[width=0.7\linewidth]{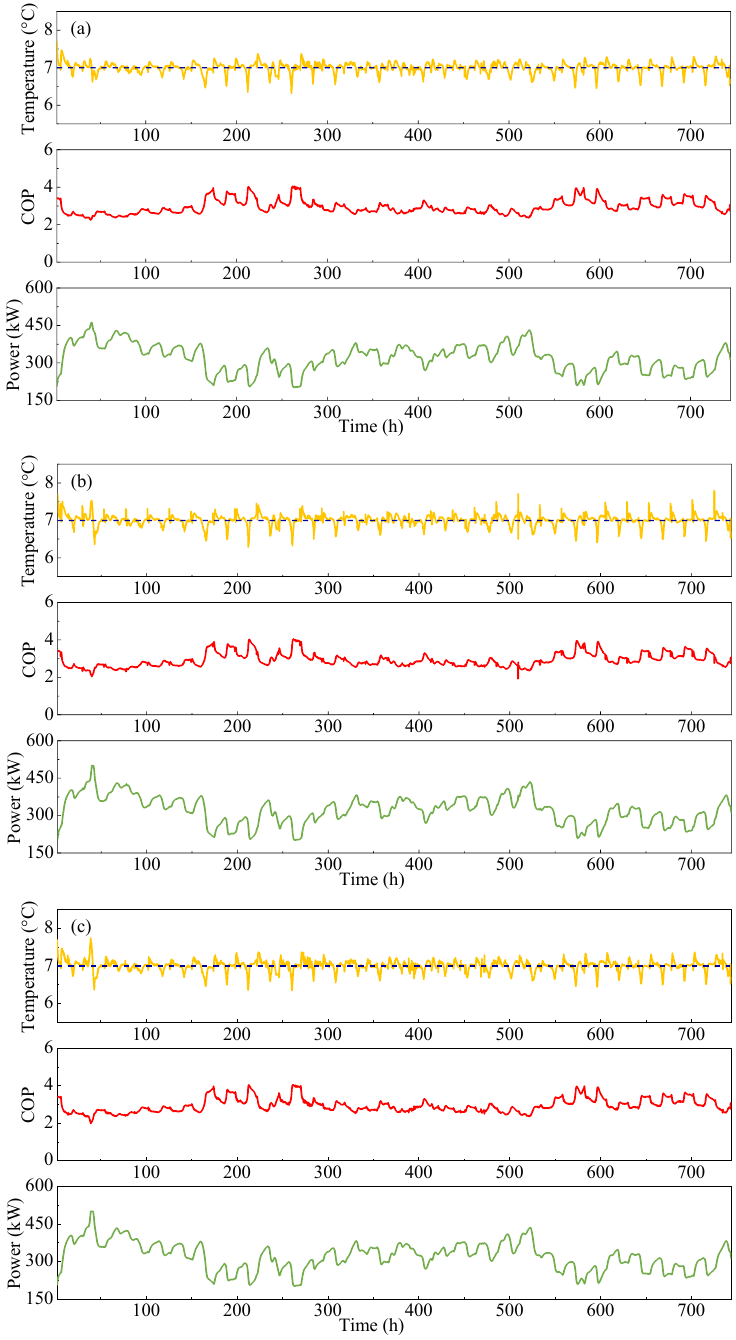}
    \caption{Evaporator outlet temperature, COP, and power use of the HP for (a) the baseline system, (b) the PCM-integrated district heating system under rule-based control, and (c) the PCM-integrated district heating system under adaptive demand-driven control.}
    \label{fig:heat_pump_performance}
\end{figure}

\figref{fig:heat_pump_performance} presents the HP performance of the baseline system and the PCM-integrated system under different control modes. The evaporator outlet temperature remains at approximately 7.0~$^{\circ}$C, with a maximum deviation within 1.0~$^{\circ}$C across all cases. The PCM-integrated case exhibits slightly greater fluctuations, which can be attributed to variations in the HP inlet temperature induced by the PCM operation. The compressor power use shows a slight increase, likely due to rises in the condenser inlet temperature after PCM integration. The average COP decreases by only about 0.67\% and 0.2\% under RBC and ADD control modes, respectively. The corresponding power use increases by merely 1.14\% and 0.39\%, compared with the baseline system. Despite these variations, all cases exhibited an effective heat recovery performance of the HP in the data center, indicating that the integration of PCM has a negligible effect.

\subsubsection{Peak-shaving and operating cost under different control strategies}\label{subsec:control_comparison}

\begin{figure}[pos=htbp]
    \centering
    \includegraphics[width=0.8\linewidth]{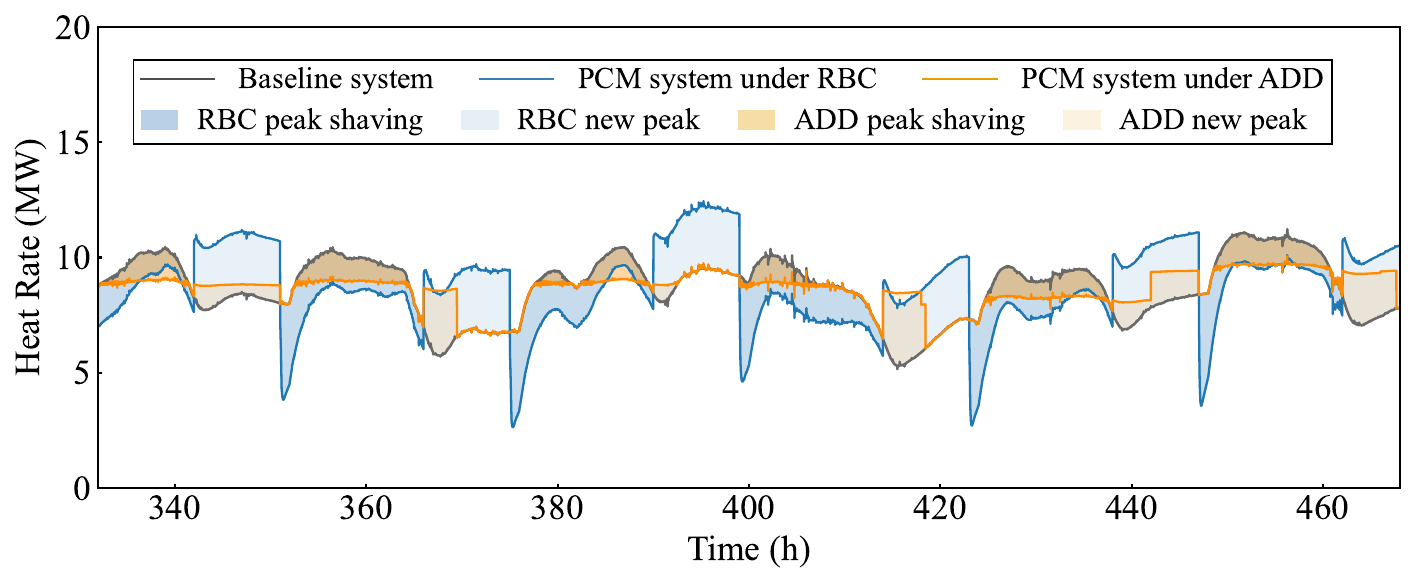}
    \caption{Heat rate supplied from the district heating network of the baseline system, the PCM-integrated system under rule-based control, and the PCM-integrated system under adaptive demand-driven control.}
    \label{fig:heat_power_comparison}
\end{figure}

After confirming that the tested cases maintain comparable thermal comfort conditions, the peak shaving and operating cost performance of the different control strategies is compared. \figref{fig:heat_power_comparison} illustrates the heat rate supplied by the DH network of the baseline system and PCM-integrated systems under different control strategies during a selected representative period. In the baseline system, relatively even peaks can be observed during each high-demand period. With the PCM-integrated system under the RBC control mode, the peak load at the beginning can be effectively reduced through PCM discharging. However, this control strategy mainly acts during the initial stage of the peak period, as the PCM storage releases heat rapidly. Consequently, the peak load reappears during the later stage of the peak period due to the lower discharging capacity, indicating limited flexibility in load management. In addition, a new load peak is observed during the off-peak period due to the high charging demand of the PCM storage tank. Therefore, although the RBC strategy demonstrates a certain capability for peak shifting, its effectiveness in overall peak reduction remains limited. In contrast, the ADD control strategy achieves a smoother reduction in the DH heat rate demand and significantly flattens the load profile over the entire operating period. The charging process is also better regulated, with the charging peak maintained below the discharging peak, thereby avoiding the occurrence of significant new peaks during off-peak periods. The controller avoids charging when the DH load approaches the threshold and increases charging only when sufficient network headroom is available. During peak-load periods, the discharge command follows the overload level while being constrained by storage availability and thermal feasibility. This operating behavior makes the ADD strategy avoid creating new off-peak peaks while maintaining discharge capability during high-demand periods. As a result, the ADD strategy provides enhanced load balancing performance and more effective peak-load mitigation compared with the RBC strategy. Meanwhile, the PCM-integrated system maintains a comparable indoor thermal comfort level, demonstrating that peak-load shifting is achieved without compromising occupant comfort.

\begin{figure}[pos=htbp]
    \centering
    \includegraphics[width=0.75\linewidth]{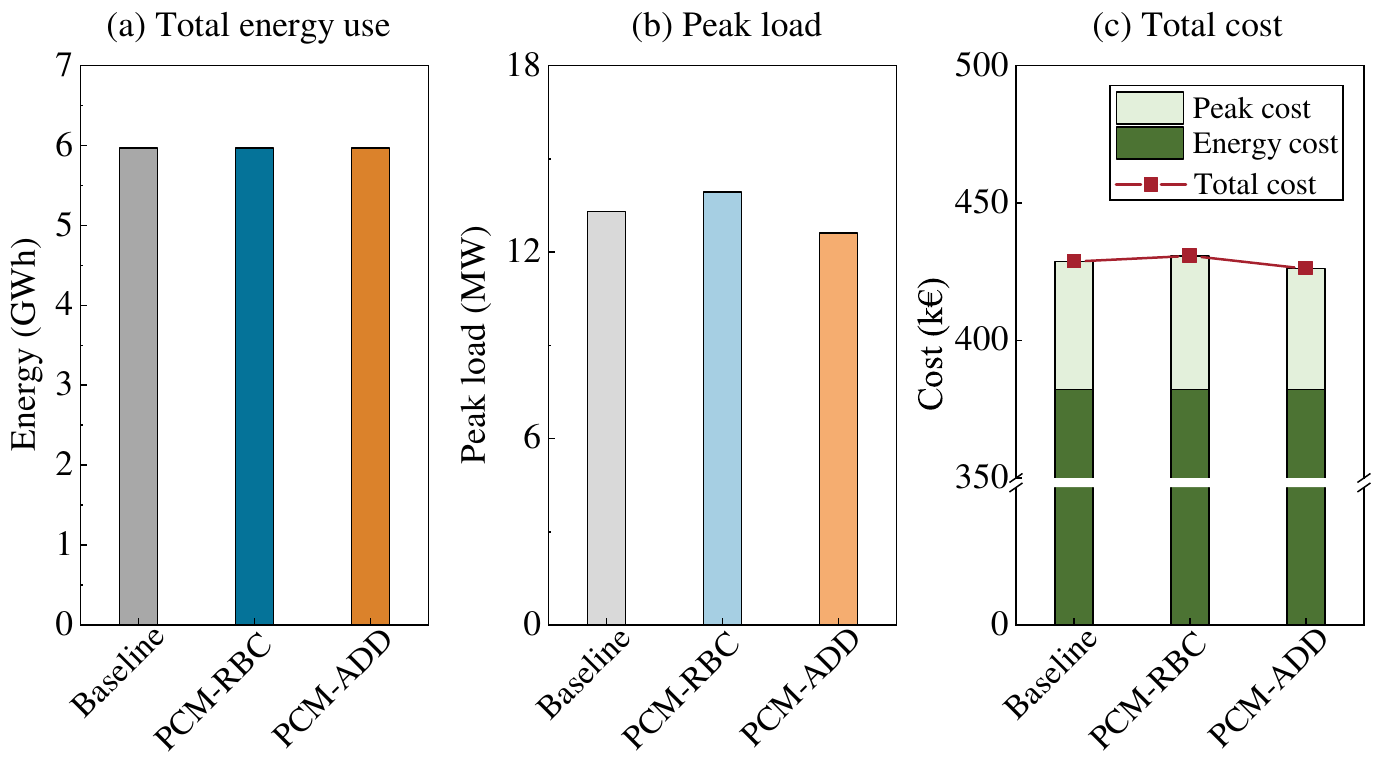}
    \caption{Comparison of district heating performance between the baseline system and the PCM-integrated system: (a) total energy use, (b) peak load, and (c) cost.}
    \label{fig:energy_peak_cost_comparison}
\end{figure}

\figref{fig:energy_peak_cost_comparison} compares the total energy use and cost of the baseline system and the PCM-integrated system under two control strategies. The results indicate that the energy cost of the DH system remains nearly unchanged across the baseline configuration and the systems integrated with PCM under both control modes. However, notable differences are observed in peak-load behavior. When the PCM-integrated system operates under RBC, the peak load increases by approximately 4.7\%, primarily due to the emergence of new charging peaks. This outcome suggests that simply integrating a PCM storage tank without advanced control may adversely affect peak demand. In contrast, the application of ADD control strategy reduces the peak load by around 5.3\%. These findings highlight that, with appropriate control, PCM integration can be flexibly implemented in DH systems and can achieve effective peak-shaving benefits.

The corresponding total cost of the DH system follows a similar trend to the peak load. Specifically, the total cost increases by approximately 0.50\%, reflecting the higher system peak load of the PCM-integrated system operating under RBC control mode. Conversely, when the peak load is reduced in the PCM-integrated system operating under the ADD control mode, the total  cost decreases by about 0.58\% compared with the baseline system. This correlation further underscores the importance of effective peak-load management in optimizing the economic performance of the system. 

The RBC case shows that a storage system can reduce part of the daytime demand but still increase the peak load if charging is not coordinated with the available network headroom. The ADD control framework enables decoupling of thermal supply from demand through strategic charging and discharging of the PCM storage, thereby improving system resilience and enabling load shifting. Therefore, the key contribution of the proposed ADD strategy is not only to release stored heat during high-demand periods, but also to prevent charging from creating new peaks. This finding suggests that the control layer is as important as the system design for demand-side peak shaving in DH networks.

\subsection{Sensitivity analysis of PCM thermophysical properties for ADD control case}\label{subsec:material_sensitivity}

The ADD control strategy offers a promising approach to reducing peak load and improving economic performance in PCM-integrated DH applications. An appropriate selection of PCM properties, including phase-change temperature,  latent heat, and thermal conductivity was further investigated under the ADD control strategy to improve the operational flexibility. %Table~\ref{tab:sensitivity_cases} lists the anaylyzed scenarios in the sensitivity analysis.

%\begin{table}[htbp]
%\centering
%\caption{List of analyzed scenarios in the sensitivity analysis.}
%\label{tab:sensitivity_cases}
%\renewcommand{\arraystretch}{0.75}
%\begin{tabular}{ccccc}
%\toprule
%Set & 
%Phase-change temperature (\si{\degreeCelsius}) &
%Latent heat (kJ/kg) &
%Thermal conductivity (W/(m$\cdot$K))
% \\
%\midrule

%\multirow{5}{*}{I}
%& 50 & \multirow{5}{*}{265} & \multirow{5}{*}{0.25} \\
%& 60 & & \\
%& 70 & & \\
%& 80 & & \\
%& 90 & & \\

%\midrule

%\multirow{5}{*}{II}
%& \multirow{5}{*}{80} & 215 & \multirow{5}{*}{0.25} \\
%& & 240 & \\
%& & 265 & \\
%& & 290 & \\
%& & 315 & \\

%\midrule

%\multirow{5}{*}{III}
%& \multirow{5}{*}{80} & \multirow{5}{*}{265} & 0.25 \\
%& & & 1.25 \\
%& & & 2.25 \\
%& & & 3.25 \\
%& & & 4.25 \\

%\bottomrule
%\end{tabular}
%\end{table}

\subsubsection{Phase-change temperature}\label{subsubsec:sens_tpcm}

\begin{figure}[pos=htbp]
    \centering
    \includegraphics[width=0.85\linewidth]{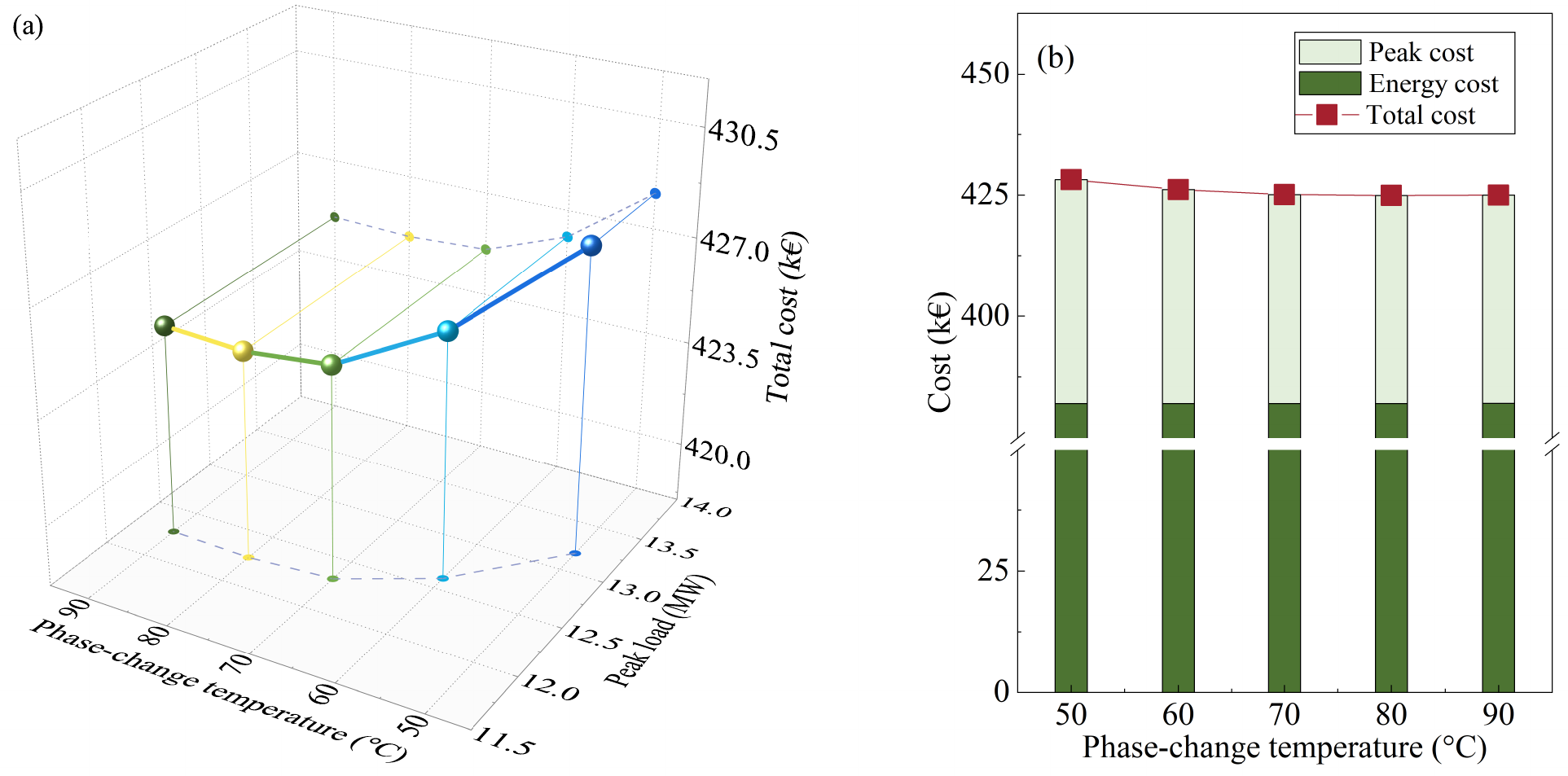}
    \caption{Peak load and operating cost of PCM-integrated systems employing PCMs with different phase-change temperatures under adaptive demand-driven control.}
    \label{fig:sens_tpcm}
\end{figure}

\figref{fig:sens_tpcm} compares the energy and cost performance of the PCM-integrated system employing PCMs with different phase-change temperatures under ADD control. The peak load and the corresponding operational cost varied noticeably with the PCM phase-change temperature. As the phase-change temperature increased from 50~$^{\circ}$C to 80~$^{\circ}$C, both the peak load and the total cost decreased simultaneously. Specifically, the peak load decreased from 13.19~MW to 12.26~MW, corresponding to an overall reduction of approximately 7.1\%, while the total cost declined from 428.14~k\texteuro{} to 424.92~k\texteuro{}, representing a reduction of about 0.8\%. Both indicators reached their minimum values at 80~$^{\circ}$C, indicating an improvement in the overall system performance and load-shifting capability. However, when the phase-change temperature was further increased to 90~$^{\circ}$C, a slight increase in both the peak load and the total cost was observed. As the PCM phase-change temperature increases beyond the effective operating range of the system, the latent heat utilization may gradually decrease due to incomplete phase-transition processes. 

\begin{figure}[pos=htbp]
    \centering
    \includegraphics[width=0.8\linewidth]{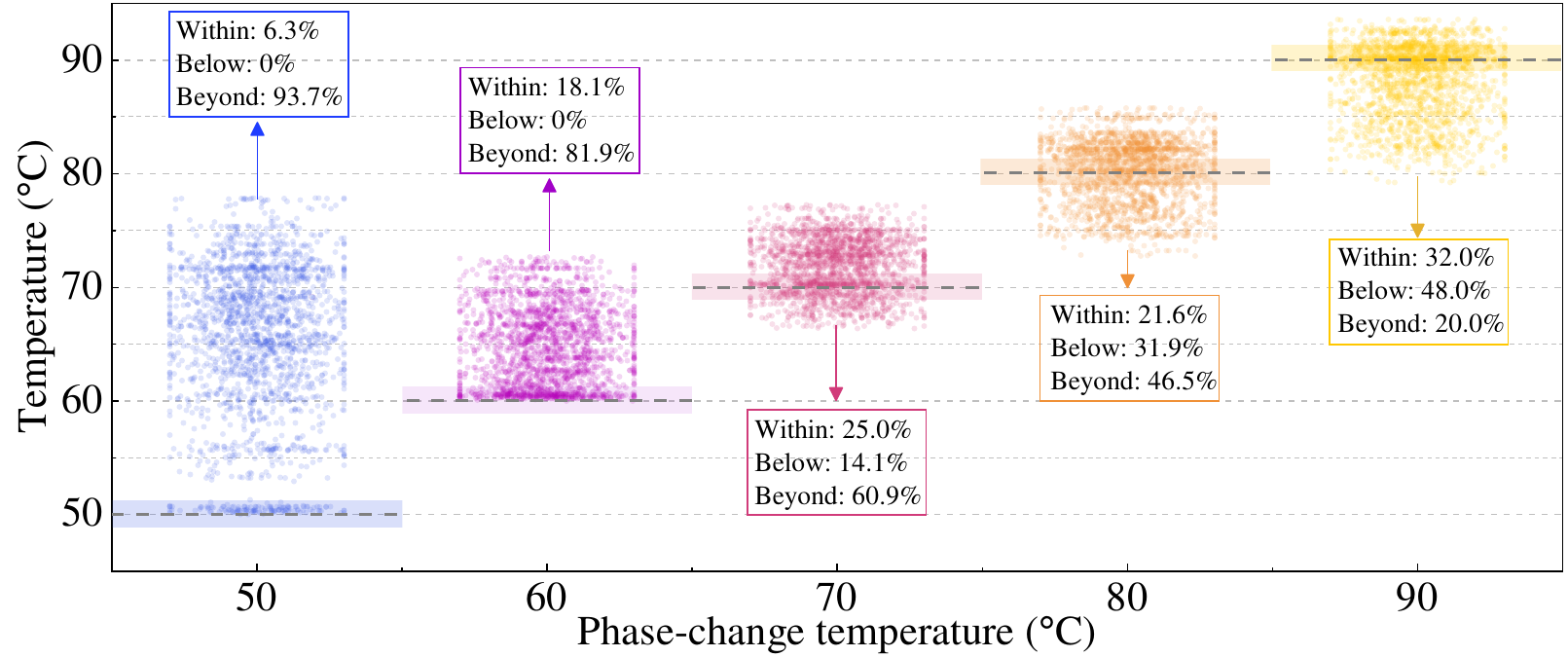}
    \caption{Average PCM temperature distribution of the storage unit employing PCMs with different phase-change temperatures.}
    \label{fig:temp_distribution_tpcm}
\end{figure}

As seen in \figref{fig:temp_distribution_tpcm}, the temperature distribution of the PCM varied significantly among the materials with different melting temperatures. For the PCM with a phase-change temperature of 50°C, approximately 90\% of the PCM remained in the liquid state throughout the simulation period, indicating that only a limited fraction of the latent heat storage capacity was effectively utilized. As the phase-change temperature increased, the proportion of PCM remaining fully melted gradually decreased, while a larger fraction of the material operated within the phase-change temperature range. For the PCM with a phase-change temperature of 70°C, nearly 85\% of the PCM temperature distribution was maintained within or beyond the phase-change band, suggesting that most of the storage material actively underwent melting and solidification processes and therefore effectively contributed to latent heat storage and release. However, when the phase-change temperature was further increased to 90°C, the effective latent heat utilization decreased because the limited temperature difference and heat transfer rate between the HTF and PCM restricted melting of the storage material, thereby reducing the fraction of PCM operating within the phase-change temperature range \cite{WuFan2024}. These results indicate that an appropriate phase-change temperature is essential for maximizing latent heat utilization, and a phase-change temperature around 70–80°C provides a better match with the investigated DH system operating conditions.

\subsubsection{Latent heat}\label{subsubsec:sens_latent}

\begin{figure}[pos=htbp]
    \centering
    \includegraphics[width=0.85\linewidth]{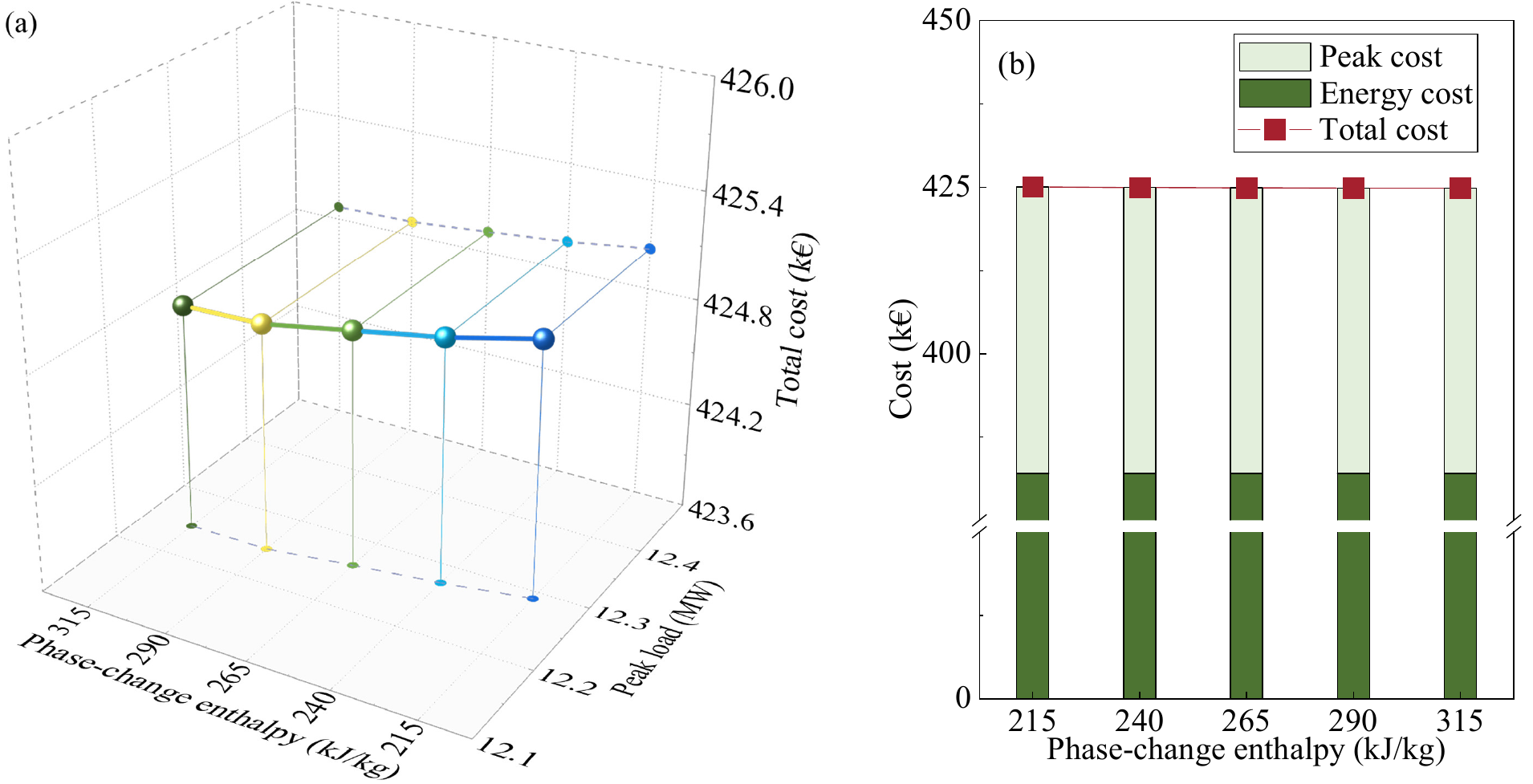}
    \caption{Peak load and total energy cost of PCM-integrated systems employing PCMs with different latent heats under adaptive demand-driven control.}
    \label{fig:sens_latent}
\end{figure}

\figref{fig:sens_latent} shows the peak-load shifting and total cost of the DH system using PCMs with different phase-change enthalpies. As the enthalpy increased, the PCM was able to store and release a larger amount of thermal energy during the charging and discharging processes, thereby enhancing the peak-shaving capability. However, within the investigated enthalpy range, only minor variations in both peak-shaving performance and economic benefits were observed. The peak-load shifting rate showed a slight increase of 0.3\% when the enthalpy increased beyond 265~kJ/kg, and the corresponding reduction in total system cost was negligible. This indicates that further increment of PCM enthalpy does not necessarily enhance system-level benefits. The limited improvement can be attributed to system operational constraints including charging and discharging schedules, heat demand characteristics, and storage utilization limits. Therefore, with a certain phase-change temperature and thermal conductivity, increasing the PCM enthalpy within the investigated range provides only slight improvements in overall DH system performance. 

\begin{figure}[pos=htbp]
    \centering
    \includegraphics[width=0.8\linewidth]{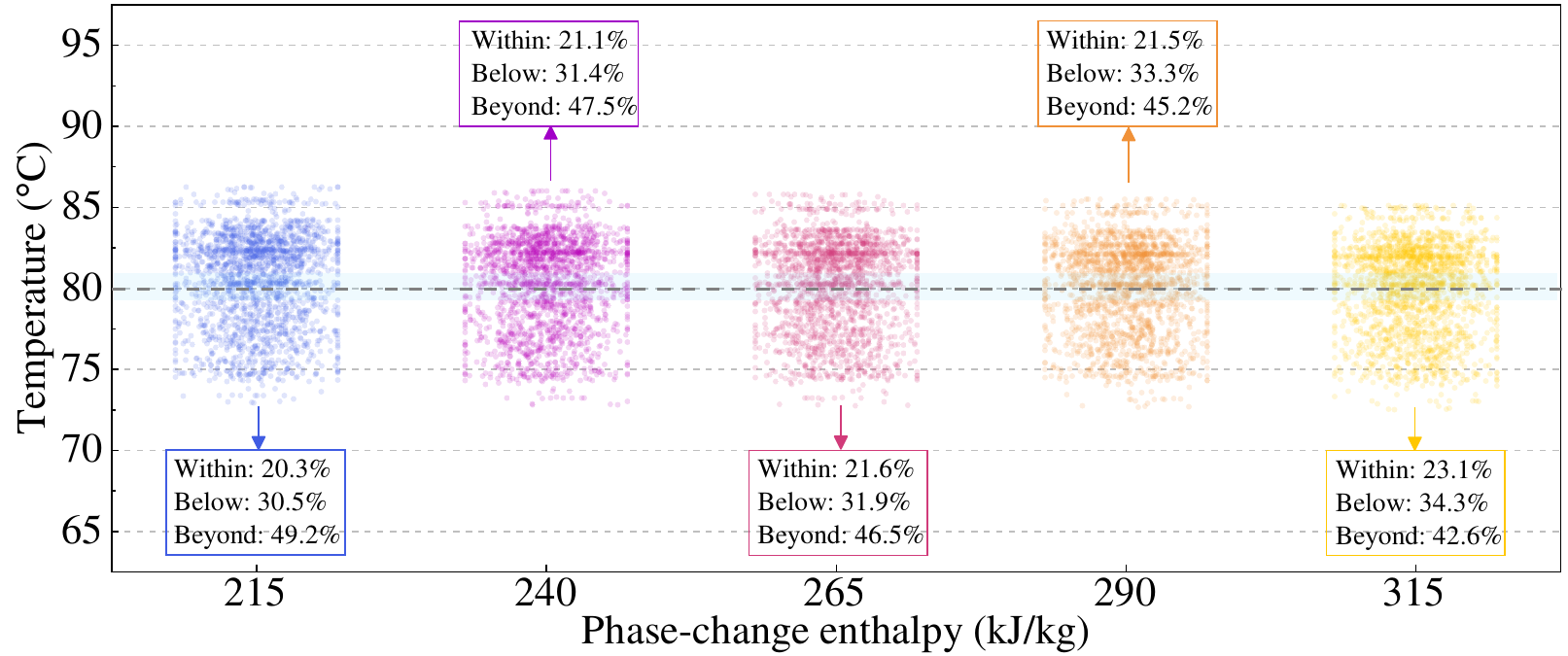}
    \caption{Average PCM temperature distribution of the storage unit employing PCMs with different phase-change enthalpies.}
    \label{fig:temp_distribution_latent}
\end{figure}

The corresponding temperature distribution of PCMs with different latent heat capacities is shown in \figref{fig:temp_distribution_latent}. As the latent heat increased, the fraction of PCM operating within the phase-change band increased slightly, indicating a minor variation in the utilization of latent heat storage. It can also be observed that the proportion of PCM temperatures beyond the phase-change range gradually decreased with increasing phase-change enthalpy. This suggests that more energy was required for the PCM to fully transition into the liquid phase, causing a larger fraction of the material to remain below or within the phase-change temperature interval. For the given phase-change temperature and thermal conductivity, the charging and discharging behavior of the PCM was primarily governed by the heat transfer process rather than the latent heat capacity. Consequently, the differences in temperature distribution among the investigated latent heat values were relatively small, resulting in comparatively slight variations in peak load and operating cost.

\subsubsection{Thermal conductivity}\label{subsubsec:sens_kpcm}

\begin{figure}[pos=htbp]
    \centering
    \includegraphics[width=0.85\linewidth]{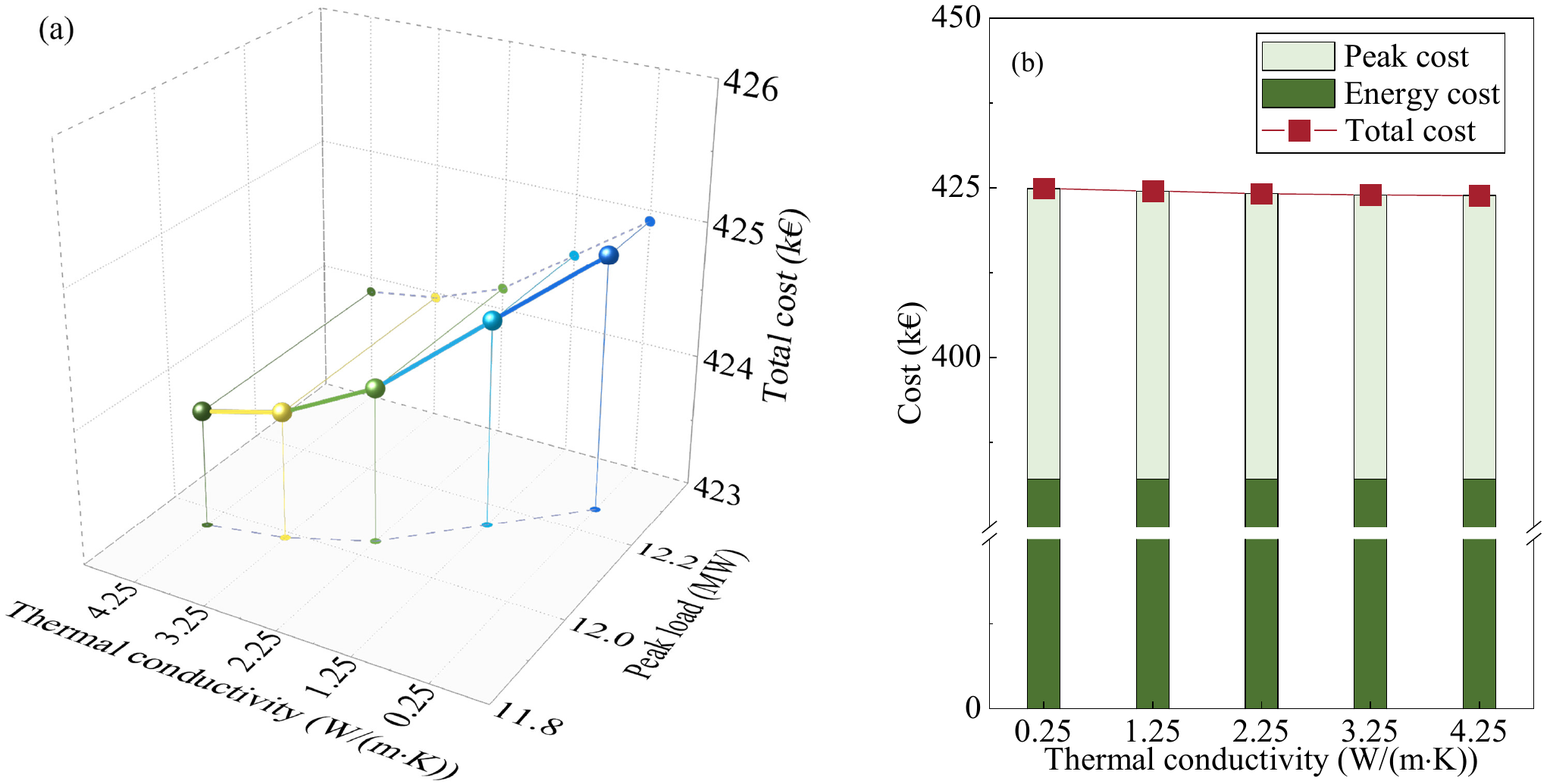}
    \caption{Peak load and total energy cost of PCM-integrated systems employing PCMs with different thermal conductivities under adaptive demand-driven control.}
    \label{fig:sens_kpcm}
\end{figure}

\figref{fig:sens_kpcm} illustrates the peak load and total cost of the PCM-integrated system using PCMs with different thermal conductivities under the ADD control strategy. The results indicate that the peak load gradually decreases with increasing thermal conductivity. An apparent reduction in peak load can be observed when the thermal conductivity increases from 0.25 to 2.25~W/(m$\cdot$K), resulting in an approximately 21.1\% larger peak load shifting rate. The improved performance at higher thermal conductivity is mainly due to enhanced heat transfer within the PCM, which enables more effective charging and discharging of the storage \cite{Yan2025CostGeometryPCM}. However, further increases in thermal conductivity provide limited additional improvement in peak-shaving performance, suggesting that internal heat transfer is no longer the primary limitation and the system becomes constrained by storage capacity and operational conditions. A similar trend can be observed for the total cost. The total cost decreases from 424.92~k\texteuro{} to 424.16~k\texteuro{} as the thermal conductivity increases from 0.25 to 2.25~W/(m$\cdot$K). Beyond this point, the total cost only exhibits a slight decrease of approximately 0.07\% under further increases in thermal conductivity. 

\begin{figure}[pos=htbp]
    \centering
    \includegraphics[width=0.8\linewidth]{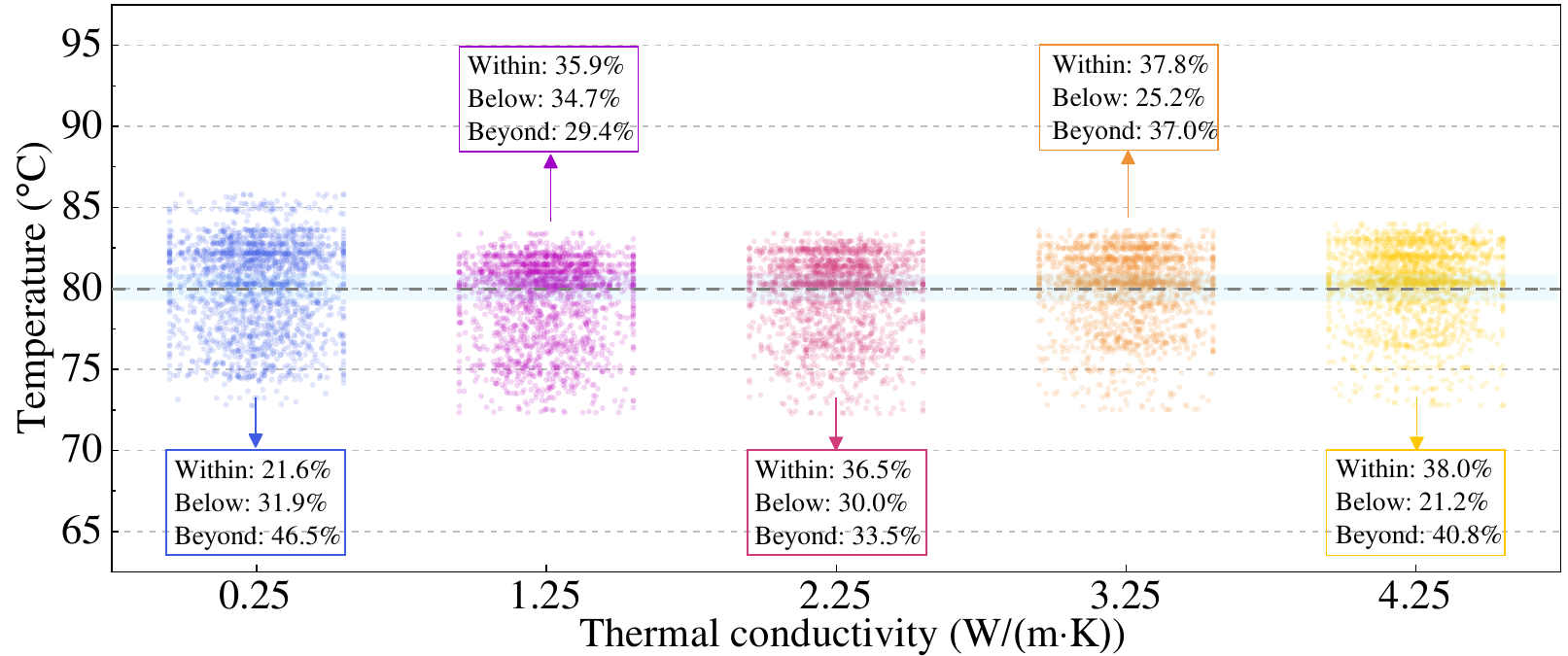}
    \caption{Average PCM temperature distribution of the storage unit employing PCMs with different thermal conductivities.}
    \label{fig:temp_distribution_kpcm}
\end{figure}

The corresponding temperature distributions of PCMs with different thermal conductivities are shown in \figref{fig:temp_distribution_kpcm}. At low thermal conductivity, limited internal heat conduction resulted in non-uniform temperature distributions, with higher temperatures near the HTF and lower temperatures in the PCM interior. Increasing the thermal conductivity to 1.25~W/(m$\cdot$K) enhanced heat transfer and reduced temperature gradients, leading to a more concentrated temperature distribution. With further increases in thermal conductivity, an increasing proportion of the PCM operated within or above the phase-change temperature range, indicating that more material could effectively participate in the melting and solidification processes \cite{Jayabal2026}. This enhanced heat transfer capability promoted more effective utilization of latent heat storage, allowing the PCM system to absorb and release thermal energy more efficiently. As a result, higher thermal conductivity contributed to improved peak-load reduction and greater operating cost savings. 

The results highlight that the phase-change temperature of the PCM should be selected to match the operating temperature range of the DH network, thereby maximizing the utilization of latent heat during both charging and discharging processes. Similarly, the thermal conductivity and latent heat of the PCM should be optimized according to the thermal and operational requirements of the system. Higher thermal conductivity can accelerate heat transfer and improve the charging and discharging rates, while higher latent heat can increase the thermal storage capacity and enhance peak-load shifting capability. However, once sufficient heat transfer performance and storage capacity have been achieved, further increases provide only limited additional improvements in peak-load reduction and economic performance. These findings suggest that PCM selection should focus on achieving an appropriate balance among phase-change temperature, latent heat, and thermal conductivity, rather than simply maximizing individual material properties.

\subsection{Sensitivity analysis of economic performance for ADD control case}
\label{subsec:economic_results}

Table~\ref{tab:economic} summarizes the annualized technical and economic performance of the baseline DH system and the proposed PCM-integrated DH system under ADD control. The integration of PCM storage tank combined with the ADD strategy significantly improved system flexibility and reduced both peak heating demand and annual operating costs. Compared with the baseline case, the ADD-controlled system achieved an annual peak-load reduction of up to 11.3\% and reduced annual operating costs by approximately 1.4\%. These improvements demonstrate the effectiveness of the proposed control strategy in coordinating PCM charging and discharging according to system demand. Despite the enhanced operational performance, the additional investment associated with PCM storage resulted in a relatively long SPP of 25.3 years. The NPV results further indicate that the current economic attractiveness of the system remains limited under existing tariff structures. Nevertheless, the achieved operational savings and improved load flexibility highlight the potential of PCM-integrated DH systems for supporting low-carbon energy systems. 

\begin{table}[htbp]
\centering
\caption{Annualized technical and economic performance of the proposed PCM-integrated district heating system with ADD control.}
\label{tab:economic}
\begin{tabular}{lcc}
\hline
Parameter & Baseline & PCM + ADD \\
\hline
Annual peak load (MW) & 98.3 & 87.2 \\
Peak load shifting rate (\%) & -- & 11.3 \\
Annual operation cost (k\euro/year) & 2892.9 & 2853.7 \\
Operating cost reduction (\%) & -- & 1.4 \\
Simple payback period (years) & -- & 25.3 \\
Net present value (NPV, k\euro) & -- & 123.7 \\
\hline
\end{tabular}
\end{table}

\begin{figure}[pos=htbp]
    \centering
    \includegraphics[width=1\linewidth]{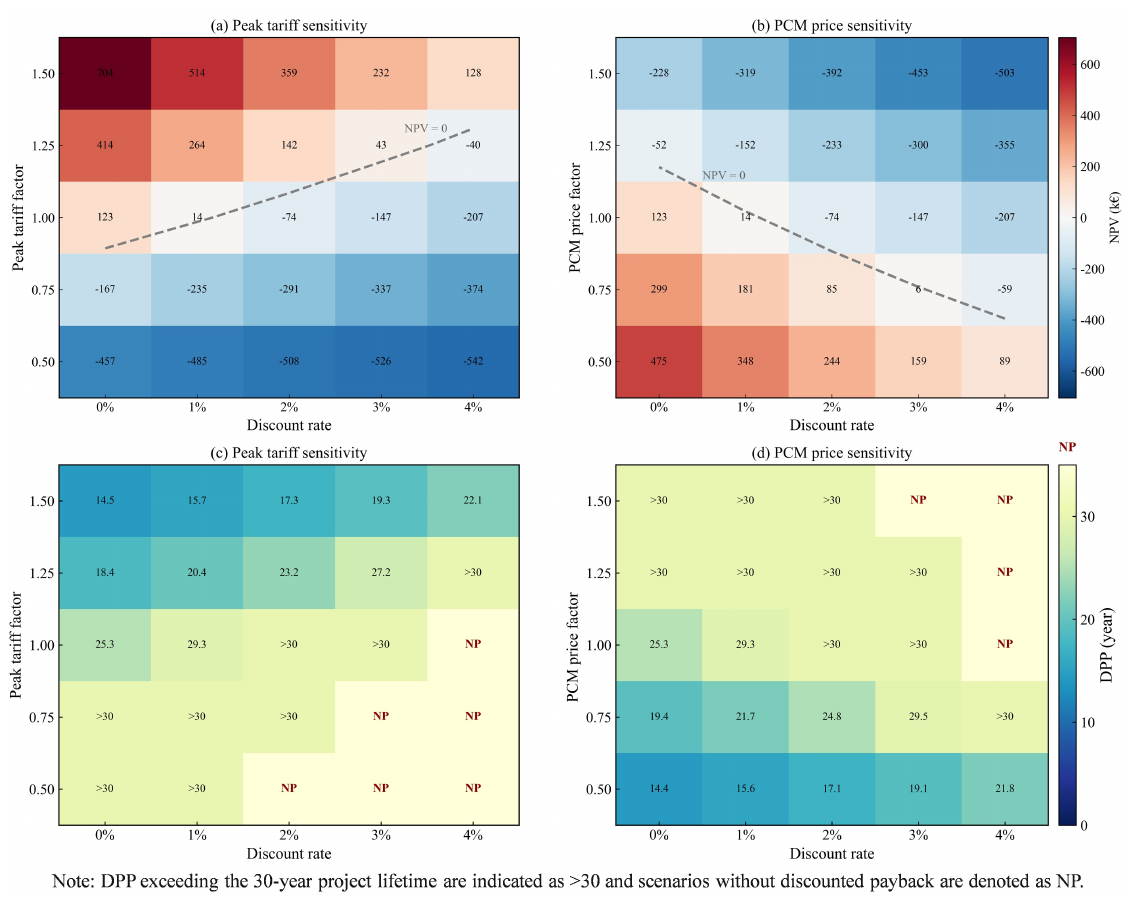}
    \caption{Sensitivity analysis of the economic performance of the PCM-integrated DH system: (a) NPV under varying peak tariff factors and discount rates; (b) NPV under varying PCM price factors and discount rates; (c) DPP under varying peak tariff factors and discount rates; and (d) DPP under varying PCM price factors and discount rates.}
    \label{fig:economic_sensitivity}
\end{figure}

Because the CAPEX and annual saving assumptions may vary with PCM price, tariff conditions, and economic environment factors, a sensitivity analysis is further conducted. As shown in \figref{fig:economic_sensitivity}(a) and (b), the NPV results indicate that both peak tariff and PCM price have substantial impacts on the economic viability of the system. Increasing the peak tariff factor significantly improves the economic performance by enhancing the value of peak-load reduction, resulting in higher NPV values and expanding the economically feasible operating region. In contrast, increasing the PCM price factor leads to a considerable reduction in NPV due to the higher CAPEX required for the PCM subsystem. As shown in \figref{fig:economic_sensitivity}(c) and (d), the payback periods calculated under different discounted rates show trends consistent with the NPV analysis. Higher peak tariffs shorten the discounted payback period (DPP) by increasing annual operational savings, whereas higher PCM prices prolong the DPP because of the increased CAPEX. The influence of the discount rate is also evident, with higher discount rates generally leading to longer DPP values and lower NPV values due to the reduced present value of future cash flows. For scenarios with high PCM prices and low peak tariffs, the DPP exceeds the project lifetime, indicating limited economic attractiveness under current market conditions. 

Although the proposed PCM-integrated DH system demonstrated significant improvements in peak-load reduction and operational flexibility, its economic viability remains constrained by the relatively high CAPEX of PCM storage system~\cite{Wu2026}. The system economic feasibility is also strongly dependent on tariff structures and financial assumptions. Further improvements in economic performance could be achieved through reductions in PCM material and installation costs and economies of scale associated with large-scale deployment \cite{LlorachMassana2017}. In addition, the current DH tariff structure primarily determines peak demand charges based on the three highest monthly demand peaks. As a result, the economic benefits associated with continuous load smoothing and demand profile optimization are not fully reflected in the current cost assessment. If future market mechanisms introduce reward-and-penalty schemes, flexibility incentives, or dynamic network tariffs that encourage smoother heat demand profiles and reduced peak fluctuations, the value of LHTES could be significantly enhanced. Under such conditions, the proposed system would not only reduce peak loads but also generate additional economic benefits, thereby substantially improving its overall economic attractiveness and accelerating its practical deployment.

Overall, the quantitative results presented in this study are associated with the investigated campus DH system, local weather conditions, building characteristics, tariff structures, and PCM storage configuration. Nevertheless, the proposed modelling and control framework is transferable to other DH systems through the adjustment of system parameters. The results provide general insights into the integration of PCM storage and ADD control for enhancing operational flexibility and peak-load management in DH networks. Future work will focus on extending the framework to incorporate additional operational uncertainties and renewable energy interactions, and further exploring advanced predictive control and optimization strategies to maximize system flexibility, economic performance, and renewable energy utilization.

\section{Conclusions}\label{sec:conclusions}

This study developed a PCM- and HP-integrated DH system to enhance energy flexibility and realize the efficient utilization of low-grade waste heat. An ADD control strategy was proposed and evaluated against a conventional RBC approach. The results showed that the ADD strategy more effectively coordinated PCM charging and discharging according to system demand, thereby avoiding the new charging peaks observed under RBC operation and resulting in a smoother load profile. Consequently, the proposed strategy achieved an annualized peak-load reduction of up to 11.3\% and an annual operating cost reduction of approximately 1.4\% compared with the baseline case, demonstrating its effectiveness in improving both load management and economic performance.

A sensitivity analysis was further conducted to systematically evaluate the impact of PCM thermophysical properties on system performance. The results indicated that phase-change temperature and thermal conductivity were the most influential parameters, while the effect of latent heat was relatively limited within the investigated range. A low phase-change temperature resulted in most of the PCM remaining in the liquid state, whereas an excessively high phase-change temperature restricted complete melting due to limited heat transfer between the HTF and PCM. Consequently, a phase-change temperature within 70--80~\si{\degreeCelsius} provided better overall performance in achieving higher peak-load reduction and cost savings. Increasing thermal conductivity enhanced heat transfer within the PCM and promoted more effective utilization of latent heat storage, leading to improved system flexibility and economic performance, although diminishing returns were observed beyond 2.25 W/(m·K). 

Despite the enhanced peak-shaving performance achieved by the proposed PCM-integrated system under ADD control strategy, the SPP remained approximately 25.3 years, indicating that further reductions in storage cost and supportive market or policy incentives are required to further enhance economic viability. Overall, the findings provide useful guidelines for PCM selection and design optimization in DH applications and the proposed ADD strategy provides an effective approach for enhancing DH flexibility, realizing waste heat recovery, and supporting the development of future low-carbon energy systems.

\section*{Acknowledgement}

This work was supported by the European Union's Horizon Europe research and innovation programme under the Marie Sk{\l}odowska-Curie Actions, grant agreement No.~101110852.

\printcredits

\bibliographystyle{elsarticle-num}
\bibliography{reference}

\end{document}